\documentclass[a4paper, USenglish, cleveref, autoref, thm-restate]{lipics-v2021}
\pdfoutput=1 %uncomment to ensure pdflatex processing (mandatatory e.g. to submit to arXiv)
\hideLIPIcs  %uncomment to remove references to LIPIcs series (logo, DOI, ...), e.g. when preparing a pre-final version to be uploaded to arXiv or another public repository
\nolinenumbers %uncomment to disable line numbering

\title{Regular Methods for Operator Precedence Languages}

\titlerunning{Regular Methods for OPLs}

\author{Thomas A. {Henzinger}}{Institute of Science and Technology Austria (ISTA), Klosterneuburg, Austria}{tah@ist.ac.at}{}{}
\author{Pavol {Kebis}}{University of Oxford, Oxford, United Kingdom}{pavol.kebis@protonmail.com}{}{}
\author{Nicolas {Mazzocchi}\footnote{Corresponding author}}{Institute of Science and Technology Austria (ISTA), Klosterneuburg, Austria \and \url{https://mazzocchi.github.io/} }{nicolas.mazzocchi@ist.ac.at}{https://orcid.org/0000-0001-6425-5369}{}
\author{N. Ege {Sara\c{c}}}{Institute of Science and Technology Austria (ISTA), Klosterneuburg, Austria}{esarac@ist.ac.at}{}{}

\authorrunning{T.\,A. Henzinger and P. Kebis and N. Mazzocchi and N.\,E. Sara\c{c}}

\Copyright{Thomas A. Henzinger and Pavol Kebis and Nicolas Mazzocchi and N. Ege {Sara\c{c}}}



\funding{This work was supported in part by the ERC-2020-AdG 101020093.}

\acknowledgements{We thank Pierre Ganty for early discussions and the anonymous reviewers for their helpful comments.}

\EventEditors{John Q. Open and Joan R. Access}
\EventNoEds{2}
\EventLongTitle{42nd Conference on Very Important Topics (CVIT 2016)}
\EventShortTitle{CVIT 2016}
\EventAcronym{CVIT}
\EventYear{2016}
\EventDate{December 24--27, 2016}
\EventLocation{Little Whinging, United Kingdom}
\EventLogo{}
\SeriesVolume{42}
\ArticleNo{23}
\usepackage[vlined, noend, linesnumbered]{algorithm2e}
\usepackage{amssymb, amsmath, bm}
\usepackage{tikz}
\usetikzlibrary{backgrounds, positioning, arrows, calc, decorations.pathmorphing}
\usepackage{mathtools}

\newcommand{\N}{\mathbb{N}}

\tikzstyle{state}=[thick,minimum size=18pt, circle,draw]
\tikzstyle{loop above right}=[out=60,in=30, min distance=5mm, looseness=8]
\tikzstyle{loop above left}=[out=150,in=120, min distance=5mm, looseness=8]
\tikzstyle{loop below left}=[out=-120,in=-150, min distance=5mm, looseness=8]
\tikzstyle{loop below right}=[out=-30,in=-60, min distance=5mm, looseness=8]
\newcommand{\angleInitialLength}{25pt}
\newcommand{\angleInitialAngle}{180}
\newcommand{\angleInitialPos}{left}
\newcommand{\angleInitialText}{}
\tikzoption{initial length}{\tikzaddafternodepathoption{\renewcommand{\angleInitialLength}{#1}}}
\tikzoption{initial angle}{\tikzaddafternodepathoption{\renewcommand{\angleInitialAngle}{#1}}}
\tikzoption{initial text}{\tikzaddafternodepathoption{\renewcommand{\angleInitialText}{#1}}}
\tikzoption{initial pos}{\tikzaddafternodepathoption{\renewcommand{\angleInitialPos}{#1}}}
\tikzstyle{initial}=[after node path={{
		[to path={[transition] (\tikztostart) --  (\tikztotarget) \tikztonodes}]
		(\tikzlastnode)++(\angleInitialAngle:\angleInitialLength) edge[at start] node[\angleInitialPos, inner sep=0pt, outer sep=1pt] {\angleInitialText} (\tikzlastnode)
}}]
\tikzstyle{transition}=[->,thick,>=stealth,shorten >=1pt,shorten <=1pt]
\tikzstyle{final}=[after node path={ node[state, scale=.8] at (\tikzlastnode) {} }]
\tikzset{nodes={font=\normalfont\normalsize\fontfamily{cmr}\selectfont}}
\renewcommand{\st}{\mid} %%%% CONFLICT <<<<
\newcommand{\stack}[1]{\ensuremath{\langle#1\rangle}}

\newcommand{\bow}{\ensuremath{\mathrel{{\triangleright}\!\!{\triangleleft}}}}
\newcommand{\shift}{%
	\ensuremath{\mathrel{%
		\mathchoice%
		{\raisebox{1pt}{$\dot{=}$}} %displaystyle
		{\raisebox{1pt}{$\dot{=}$}} %textstyle
		{\raisebox{.8pt}{\scriptsize$\dot{=}$}} %scriptstyle
		{\raisebox{.6pt}{\tiny$\dot{=}$}} %scriptscriptstyle
	}}
}
\newcommand{\nshift}{%
	\ensuremath{\mathrel{%
			\mathchoice%
			{\makebox[.8em]{\makebox[0pt]{\raisebox{1pt}{$\neq$}}{\makebox[0pt]{\raisebox{1pt}{$\dot{=}$}}}}} %displaystyle
			{\makebox[.8em]{\makebox[0pt]{\raisebox{1pt}{$\neq$}}{\makebox[0pt]{\raisebox{1pt}{$\dot{=}$}}}}} %textstyle
			{\makebox[.8em]{\makebox[0pt]{\raisebox{.8pt}{\scriptsize$\neq$}}{\makebox[0pt]{\raisebox{.8pt}{\scriptsize$\dot{=}$}}}}} %scriptstyle
			{\makebox[.8em]{\makebox[0pt]{\raisebox{.6pt}{{\tiny$\neq$}}}{\makebox[0pt]{\raisebox{.6pt}{\tiny$\dot{=}$}}}}} %scriptscriptstyle
	}}
}

\newcommand{\push}{%
	\ensuremath{\mathrel{%
			\mathchoice%
			{\raisebox{1pt}{$\lessdot$}} %displaystyle
			{\raisebox{1pt}{$\lessdot$}} %textstyle
			{\raisebox{.8pt}{\scriptsize$\lessdot$}} %scriptstyle
			{\raisebox{.6pt}{\tiny$\lessdot$}} %scriptscriptstyle
	}}
}
\newcommand{\pusheq}{%
	\ensuremath{\mathrel{%
			\mathchoice%
			{\makebox[.8em]{{\makebox[0pt]{$\leq$}}{\makebox[0pt]{\raisebox{1pt}{$\lessdot$}}}}} %displaystyle
			{\makebox[.8em]{{\makebox[0pt]{$\leq$}}{\makebox[0pt]{\raisebox{1pt}{$\lessdot$}}}}} %textstyle
			{\makebox[.8em]{{\makebox[0pt]{\scriptsize$\leq$}}{\makebox[0pt]{\raisebox{.8pt}{\scriptsize$\lessdot$}}}}} %scriptstyle
			{\makebox[.8em]{{\makebox[0pt]{\tiny$\leq$}}{\makebox[0pt]{\raisebox{.6pt}{\tiny$\lessdot$}}}}} %scriptscriptstyle
	}}
}
\newcommand{\pop}{%
	\ensuremath{\mathrel{%
			\mathchoice%
			{\raisebox{1pt}{$\gtrdot$}} %displaystyle
			{\raisebox{1pt}{$\gtrdot$}} %textstyle
			{\raisebox{.8pt}{\scriptsize$\gtrdot$}} %scriptstyle
			{\raisebox{.6pt}{\tiny$\gtrdot$}} %scriptscriptstyle
	}}
}
\newcommand{\popeq}{%
	\ensuremath{\mathrel{%
			\mathchoice%
			{\makebox[.8em]{{\makebox[0pt]{$\geq$}}{\makebox[0pt]{\raisebox{1pt}{$\gtrdot$}}}}} %displaystyle
			{\makebox[.8em]{{\makebox[0pt]{$\geq$}}{\makebox[0pt]{\raisebox{1pt}{$\gtrdot$}}}}} %textstyle
			{\makebox[.8em]{{\makebox[0pt]{\scriptsize$\geq$}}{\makebox[0pt]{\raisebox{.8pt}{\scriptsize$\gtrdot$}}}}} %scriptstyle
			{\makebox[.8em]{{\makebox[0pt]{\tiny$\geq$}}{\makebox[0pt]{\raisebox{.6pt}{\tiny$\gtrdot$}}}}} %scriptscriptstyle
	}}
}
\newcommand{\chain}[3]{\bgroup\ensuremath{^{#1}[#2]^{#3}}\egroup}
\newcommand{\first}{\ensuremath{\triangleleft}}
\newcommand{\last}{\ensuremath{\triangleright}}

\newcommand{\A}{\mathcal{A}\xspace}
\newcommand{\B}{\mathcal{B}\xspace}
\newcommand{\C}{\mathcal{C}\xspace}

\newcommand{\tikzpush}[1]{\mathrel{\raisebox{-.4pt}[0pt][0pt]{\tikz[anchor=center] \draw[-stealth] (0mm,0mm) -- node[above] {\scriptsize$#1$}(4mm,0mm);}}}
\newcommand{\tikzshift}[1]{\mathrel{\raisebox{-.4pt}[0pt][0pt]{\tikz[anchor=center] \draw[-stealth, dashed] (0mm,0mm) -- node[above] {\scriptsize$#1$}(4mm,0mm);}}}
\newcommand{\tikzpop}[1]{\mathrel{\raisebox{-.4pt}[0pt][0pt]{\tikz[anchor=center] \draw[double, -stealth] (0mm,0mm) -- node[above] {\scriptsize$#1$}(4mm,0mm);}}}
\newcommand{\tikztransition}{\mathrel{\raisebox{0pt}[0pt][0pt]{\tikz[anchor=center] \draw[-stealth, decorate,decoration={snake,amplitude=.4mm,segment length=1.3mm,post length=.8mm}] (0mm,0mm) -- (4mm,0mm);}}}
\newcommand{\tikzntransition}{\mathrel{\raisebox{-2pt}[0pt][0pt]{\tikz[anchor=center] \draw[-stealth, decorate,decoration={snake,amplitude=.4mm,segment length=1.3mm,post length=.8mm}] (0mm,0mm) -- node {\raisebox{-.5pt}[0pt][0pt]{\scriptsize$/$}} (4mm,0mm);}}}
\newcommand{\tikzpath}[1][\empty]{\ifx#1\empty\tikztransition^*\else\tikztransition^{#1}\fi}
\newcommand{\tikznpath}[1][\empty]{\ifx#1\empty\tikzntransition^*\else\tikzntransition^{#1}\fi}

\begin{document}
	\maketitle
	\begin{abstract}
		The operator precedence languages (OPLs) represent the largest known subclass of the context-free languages which enjoys all desirable closure and decidability properties.
		This includes the decidability of language inclusion, which is the ultimate verification problem.
		Operator precedence grammars, automata, and logics have been investigated and used, for example, to verify programs with arithmetic expressions and exceptions (both of which are deterministic pushdown but lie outside the scope of the visibly pushdown languages).
		In this paper, we complete the picture and give, for the first time, an algebraic characterization of the class of OPLs in the form of a syntactic congruence that has finitely many equivalence classes exactly for the operator precedence languages.
		This is a generalization of the celebrated Myhill-Nerode theorem for the regular languages to OPLs.
		As one of the consequences, we show that universality and language inclusion for nondeterministic operator precedence automata can be solved by an antichain algorithm.
		Antichain algorithms avoid determinization and complementation through an explicit subset construction, by leveraging a quasi-order on words, which allows the pruning of the search space for counterexample words without sacrificing completeness.
		Antichain algorithms can be implemented symbolically, and these implementations are today the best-performing algorithms in practice for the inclusion of finite automata.
		We give a generic construction of the quasi-order needed for antichain algorithms from a finite syntactic congruence.
		This yields the first antichain algorithm for OPLs, an algorithm that solves the \textsc{ExpTime}-hard language inclusion problem for OPLs in exponential time.
	\end{abstract}

	\section{Introduction}

	Pushdown automata are a fundamental model of computation and the preferred formalism to parse programs in a deterministic manner.
	In verification, they are used to encode the behaviors of both systems and specifications that involve, for example, nested procedure calls.
	However, unlike for regular languages specified by finite automata, the inclusion of context-free languages given by pushdown automata is undecidable, even for deterministic machines.
	This is why expressive subclasses of context-free languages with decidable properties have been studied in the past decades.
	Prominent among those formalisms is the class of visibly pushdown languages~\cite{DBLP:conf/stoc/AlurM04}, which is strictly contained in the deterministic context-free languages.
	A visibly pushdown language (VPL) is a context-free language where each word admits a single parse tree, which does not depend on the pushdown automaton that generates (or accepts) the word.
	More technically, visibly pushdown automata (VPDAs) extend finite automata with a memory stack that is restricted to ``push'' and ``pop'' operations on disjoint subsets of the input alphabet.
	VPDAs have become popular in verification for several reasons.
	First, they recognize ``well-nested'' words, which find applications in the analysis of HTML and XML documents.
	Second, their restricted stack behavior enables desirable closure and decidability properties;
	in particular, in contrast to deterministic context-free languages, VPDAs can be complemented and their inclusion is decidable.
	Third, the VPLs admit a generalization of the celebrated Myhill-Nerode theorem for the regular languages~\cite{DBLP:conf/icalp/AlurKMV05}:
	they can be characterized algebraically by a finite syntactic congruence, which not only explains the decidability results, but also leads to symbolic verification algorithms,
	such as antichain-based universality and inclusion checking for VPDAs~\cite{DBLP:conf/lata/BruyereDG13}.
	
	There are, however, important languages that are parsable by deterministic pushdown automata, yet are not visibly pushdown.
	An important example are the arithmetic expressions with two binary operators, addition and multiplication, where multiplication takes precedence over addition.
	Most programming languages allow such expressions with implicit precedence relations between operators, instead of insisting on explicit parantheses to disambiguate.
	For this very purpose, Floyd introduced three elementary precedence relations between letters, namely, \emph{equals in precedence} $\shift$, \emph{yields precedence} $\push$, and \emph{takes precedence} $\pop$, which provide structure to words.
	He introduced the \emph{operator precedence languages} (OPLs), a subclass of the context-free languages, where non-conflicting precedence relations between letters can be derived from the context-free grammar~\cite{DBLP:journals/jacm/Floyd63}. 
	The ability to extract non-conflicting relations from the grammar provides a unique parse tree for each word. 
	However, unlike for VPLs, a letter is not assigned to a unique stack operation, but will trigger ``push'' and ``pop'' operations depending on its precedence with respect to the adjacent letters.
	This allows OPLs to model not only arithmetic expressions, but also languages with exception handling capabilities, where a single closed parenthesis may close several open parentheses~\cite{DBLP:journals/fmsd/AlurF21,DBLP:conf/sefm/PontiggiaCP21}.
	
	The class of OPLs lies strictly between the VPLs and the deterministic context-free languages.
	Despite their extra expressive power, the OPLs enjoy the closure and decidability properties of the VPLs, and they even do so at the same cost in computational complexity:
	the class of OPLs is closed under all boolean and regular operations (union, intersection, complement, concatenation, reverse, and Kleene star)~\cite{DBLP:journals/jcss/Crespi-ReghizziM12,DBLP:journals/iandc/Crespi-ReghizziMM78};
	their emptiness can be solved in \textsc{PTime} (it is \textsc{PTime}-hard for VPDAs), and universality and inclusion in \textsc{ExpTime} (they are \textsc{ExpTime}-hard for VPDAs)~\cite{DBLP:journals/siamcomp/LonatiMPP15}.
	Moreover, OPLs admit a logical characterization in terms of a monadic second-order theory over words, as well as an operational characterization in terms of automata with a stack (called OPAs)~\cite{DBLP:journals/siamcomp/LonatiMPP15}.
	In short, OPLs offer many of the benefits of the VPLs at no extra cost.
	
	In this paper, we complete the picture by showing that OPLs also offer an algebraic characterization in form of a generalized Myhill-Nerode theorem.
	Specifically, we define a syntactic congruence relation $\equiv_L$ for languages $L$ such that $\equiv_L$ has finitely many equivalence classes if and only if $L$ is an OPL.
	Finite syntactic congruences provide a formalism-independent (i.e., grammar- and automaton-independent) definition for capturing the algebraic essence of a class of languages.
	In addition to the regular languages (Myhill-Nerode) and the VPLs, such congruences have been given	also for tree languages~\cite{DBLP:journals/ijfcs/Gecseg07},
	for profinite languages~\cite{DBLP:conf/stacs/Pin09},
	for omega-regular languages~\cite{DBLP:journals/tcs/Arnold85,DBLP:journals/tcs/MalerS97},
	for sequential and rational transducers~\cite{DBLP:journals/tcs/Choffrut03,DBLP:journals/lmcs/FiliotGL19}.
	Furthermore, such characterization results through syntactic congruences have been used to design determinization~\cite{DBLP:conf/icalp/AlurKMV05,DBLP:journals/isci/IgnjatovicCB08}, minimization~\cite{DBLP:conf/mfcs/GantyGV19,DBLP:conf/concur/KumarMV06}, and learning~\cite{DBLP:conf/tacas/BruyerePS22,DBLP:conf/concur/KumarMV06,DBLP:conf/mfcs/MichaliszynO22} algorithms.
	
	Our contribution in this paper is twofold.
	Besides giving a finite congruence-based characterization of OPLs, we show how such a characterization can be used to obtain antichain-based verification algorithms,
	i.e., symbolic algorithms for checking the universality and inclusion of operator precedence automata (OPA).
	Checking language inclusion is the paradigmatic verification problem for any automaton-based specification formalism, but it is also computationally difficult:
	{\sc PSpace}-hard for finite automata, {\sc ExpTime}-hard for VPDAs, undecidable for pushdown automata.
	This is why the verification community has devised and implemented symbolic algorithms, which avoid explicit subset constructions for determinization and complementation by manipulating symbolic representations of sets of states.
	For finite automata, the antichain-based algorithms have proven to be particularly efficient in practice: DWINA~\cite{DBLP:journals/acta/FiedorHLV19} outperforms MONA~\cite{DBLP:journals/ijfcs/KlarlundMS02} for deciding WS1S formulae, ATC4VPA~\cite{DBLP:conf/lata/BruyereDG13} outperforms VPAchecker~\cite{DBLP:conf/lata/TangO12} for deciding VPDAs inclusion, and Acacia~\cite{DBLP:conf/cav/FiliotJR09} outperforms Lily~\cite{DBLP:conf/fmcad/JobstmannB06} for LTL synthesis. % the antichain tool
	%For finite automata, the antichain-based algorithms have proven to be particularly efficient in practice~\cite{DBLP:conf/cav/WulfDHR06,DBLP:conf/tacas/DoyenR10,DBLP:conf/tacas/AbdullaCHMV10}. \rednote{ege: not sure what tom wants here, he only had "?" for the references so i added the last two}
	They leverage a quasi-order on words to prune the search for counterexamples.
	Intuitively, whenever two words are candidates to contradict the inclusion between two given languages, and the words are related by the quasi-order at hand,
	the ``greater'' word can be discarded without compromising the completeness of the search.
	During symbolic fixpoint iteration, this ``quasi-order reduction'' yields a succinct representation of intermediate state sets.
	Based on our syntactic congurence, we show how to systematically compute a quasi-order that enables the antichain approach.
	Then, we provide the first antichain algorithm for checking language inclusion (and as a special case, universality) between OPAs.
	In fact, our antichain inclusion algorithm can take any suitable syntactic congruence over structured words (more precisely, any finite equivalence relation that is monotonic for structured words and saturates its language).
	The instantiation of the antichain algorithm with our syntactic congruence yields an \textsc{ExpTime} algorithm for the inclusion of OPAs, which is optimal in terms of enumeration complexity.
	
	In summary, we generalize two of the most appealing features of the regular languages---the finite characterization by a syntactic congruence, and the antichain inclusion algorithm---to 
	the important context-free subclass of operator precedence languages.

	\subparagraph*{Overview.}
	In Section~2, we define operator precedence alphabets and structured words.
	We present operator precedence grammars as originally defined by Floyd.
	We then define the operator precedence languages (OPLs) together with their automaton model (OPAs).
	Finally, we summarize the known closure and complexity results for OPLs and OPAs.
	In Section~3, we introduce the syntactic congruence that characterizes the class of OPLs.
	Subsection~3.1 proves that the syntactic congruence of every OPLs has finitely many equivalence classes,
	and Subsection~3.2 proves that every language whose syntactic congruence has finitely many equivalence classes is an OPL.
	In Section~4, we present our antichain inclusion algorithm.
	First, we introduce the notion of a language abstraction and prove that our syntactic congruence is a language abstraction of OPLs. 
	We also present a quasi-order that relaxes the syntactic congruence while preserving the property of being a language abstraction.
	Then, we provide an antichain algorithm that decides the inclusion between automata whose languages have finite abstractions. %an OPA and a language together with its abstraction.
	We prove the correctness of our algorithm and establish its complexity on OPAs.
	In Section~5, we conclude with future directions.

	\subparagraph*{Related Work.}
	Operator precedence grammars and their languages were introduced by Floyd~\cite{DBLP:journals/jacm/Floyd63} with the motivation to construct efficient parsers.
	Inspired by Floyd's work, Wirth~and~Weber~\cite{DBLP:journals/cacm/WirthW66} defined simple precedence grammars as the basis of an ALGOL-like language.
	The relation between these two models was studied in~\cite{DBLP:conf/stoc/Fischer69}.
	The properties of OPLs were studied in~\cite{DBLP:journals/siamcomp/Crespi-ReghizziGM81,DBLP:journals/iandc/Crespi-ReghizziMM78}.
	Later, their relation with the class of VPLs was established in~\cite{DBLP:journals/jcss/Crespi-ReghizziM12},
	their parallel parsing was explored in~\cite{DBLP:journals/ipl/BarenghiCMP13},
	and automata-theoretic and logical characterizations were provided in~\cite{DBLP:journals/siamcomp/LonatiMPP15}.
	Recent contributions provide a model-checking algorithm for operator precedence automata~\cite{DBLP:journals/tcs/ChiariMP20},
	a generalization to a weighted model~\cite{DBLP:journals/iandc/DrosteDMP22},
	and their application to verifying procedural programs with exceptions~\cite{DBLP:conf/sefm/PontiggiaCP21}.
	
	The OPLs form a class of structured context-free languages~\cite{DBLP:journals/csr/MandrioliP18}
	that sits strictly between deterministic context-free languages and the VPLs~\cite{DBLP:conf/stoc/AlurM04,DBLP:journals/corr/abs-0907-2130}.
	To the best of our knowledge, the OPLs constitute the largest known class that enjoys all desired closure and decidability properties.
	Several attempts have been made to move beyond this class, however, this often comes at the cost of losing some desirable property.
%	For example, the locally chain-parsable languages~\cite{DBLP:journals/tcs/Crespi-Reghizzi17} and the higher-order OPLs with fixed order~\cite{DBLP:journals/jcss/Crespi-Reghizzi20} are not closed under concatenation.
	For example, the locally chain-parsable languages are not closed under concatenation and Kleene star~\cite{DBLP:journals/tcs/Crespi-Reghizzi17},
	and the higher-order OPLs with fixed order are not closed under concatenation~\cite{DBLP:journals/jcss/Crespi-Reghizzi20}.
	Despite the fact that they are more powerful than the VPLs and enjoy all closure and decidability properties,
	the class of OPLs is not nearly as well studied.
	In particular, a finite syntactic congruence characterizing the VPLs was provided in~\cite{DBLP:conf/icalp/AlurKMV05}.
	An analogous result was missing for the OPLs until now.

	The antichain algorithm for checking language inclusion was originally introduced for finite automata~\cite{DBLP:conf/cav/WulfDHR06}
	and later extended to alternating finite automata~\cite{DBLP:conf/tacas/WulfDMR08}.
	The approach has been adapted to solve games with imperfect information~\cite{DBLP:conf/csl/ChatterjeeDHR06},
	the inclusion of tree automata~\cite{DBLP:conf/wia/BouajjaniHHTV08},
	the realizability of linear temporal logic~\cite{DBLP:conf/cav/FiliotJR09},
	the satisfiability of quantified boolean formulas~\cite{DBLP:conf/atva/BrihayeBDDR11},
	the inclusion of visibly pushdown automata~\cite{DBLP:conf/lata/BruyereDG13},
	the inclusion of $\omega$-visibly pushdown automata~\cite{doveriAntichainsAlgorithmsInclusion2023},
	the satisfiability of weak monadic second-order logic~\cite{DBLP:conf/tacas/FiedorHLV15},
	and the inclusion of B\"uchi automata~\cite{DBLP:conf/cav/DoveriGM22,DBLP:conf/concur/DoveriGPR21}.
	The antichain-based approach can be expressed as a complete abstract interpretation as it is captured by the framework introduced in~\cite{DBLP:conf/sas/GantyR019,DBLP:journals/tocl/GantyRV21}.
	We provide the first antichain inclusion algorithm for OPLs,
	and the first generic method to construct an antichain algorithm from a finite syntactic congruence.

	\section{Operator Precedence Languages}
	We assume that the reader is familiar with formal language theory.
	
	\subsection{Operator Precedence Relations and Structured Words}

	Let $\Sigma$ be a finite alphabet.
	We refer by $\Sigma^*$ to the set of all words over $\Sigma$, by $\varepsilon$ to the empty word, and we let $\Sigma^+ = \Sigma^* \setminus \{\varepsilon\}$.
	Given a word $w \in \Sigma^*$, we denote by $|w|$ its length, by $w^\first$ its first letter, and by $w^\last$ its last letter.
	In particular $|\varepsilon|=0$, $\varepsilon^{\first} = \varepsilon$, and $\varepsilon^{\last} = \varepsilon$.
%	We also define $\Sigma^{\leq n} = \{w \in \Sigma \st |w|\leq n\}$ for all $n\in\N$. %\rednote{not used anywhere?}

	An \emph{operator precedence alphabet} $\widehat{\Sigma}$ is an alphabet $\Sigma$ equipped with the precedence relations $\push$, $\pop$, $\shift$, given by a matrix (see \figurename~\ref{table:arithmetic}).
	Formally, for each ordered pair of letters $(a,b) \in \Sigma^2$, exactly one\footnote[1]{In the literature, operator precedence matrices are defined over sets of precedence relations, leading then to notion of precedence conflict. We use the restriction to singletons because it covers the interesting part of the theory.} of the following holds:
	\begin{itemize}
		\item \emph{$a$ yields precedence to $b$}, denoted $a \push b$,
		\item \emph{$a$ takes precedence over $b$}, denoted $a \pop b$,
		\item \emph{$a$ equals in precedence with $b$}, denoted $a \shift b$.
	\end{itemize}
	
	\newcommand{\lp}{(\!|}
	\newcommand{\rp}{|\!)}
	\begin{figure}
		\begin{minipage}{.25\linewidth}
			\centering
			\scalebox{.9}{
			\bgroup
			\renewcommand{\arraystretch}{.9}
			\setlength{\arraycolsep}{2pt}
			$\begin{array}{ c |c c c c c c c}
				&    +    & \times  &    0    &    1    &    \lp    &    \rp    &  \varepsilon  \\\hline
				+      &  \pop   &  \push  &  \push  &  \push  &  \push  &  \pop   &  \pop         \\
				\times    &  \pop   &  \pop   &  \push  &  \push  &  \push  &  \pop   &  \pop         \\
				0       &  \pop   &  \pop   &  \cdot  &  \cdot  &  \cdot  &  \pop   &  \pop         \\
				1       &  \pop   &  \pop   &  \cdot  &  \cdot  &  \cdot  &  \pop   &  \pop         \\
				\lp       &  \push  &  \push  &  \push  &  \push  &  \push  &  \shift &  \pop         \\
				\rp      &  \pop   &  \pop   &  \cdot  &  \cdot  &  \cdot  &  \pop   &  \pop         \\
				\varepsilon &  \push  &  \push  &  \push  &  \push  &  \push  &  \push  &  \shift       \\
			\end{array}$
			\egroup}
		\end{minipage}
		\begin{minipage}{.54\linewidth}
			\centering
			\scalebox{.9}{$\begin{array}{c}
				\varepsilon \push 1 \pop + \push 0 \pop \times \push \lp \push \pmb{1} \pop + \push \pmb{1} \pop \rp \pop \varepsilon \\
				\varepsilon \push 1 \pop + \push 0 \pop \times \push \lp \push \pmb{+} \pop \rp \pop \varepsilon \\
				\varepsilon \push 1 \pop + \push \pmb{0} \pop \times \push \pmb{\lp} \shift \pmb{\rp} \pop \varepsilon\\
				\varepsilon \push \pmb{1} \pop + \push \pmb{\times} \pop \varepsilon \\
				\varepsilon \push \pmb{+} \pop \varepsilon \\
				\varepsilon \shift \varepsilon \\
			\end{array}$}
		\end{minipage}
		\begin{minipage}{.15\linewidth}
			\centering
			\scalebox{.9}{$\begin{array}{c}
				1 + 0 \times \lp \pmb{1} + \pmb{1} \rp\\
				1 + 0 \times \lp A \pmb{+} B \rp\\
				1 + \pmb{0} \times \pmb{\lp} A \pmb{\rp}\\
				\pmb{1} + B \pmb{\times} C\\
				A \pmb{+} B\\
				A\\
			\end{array}$}
		\end{minipage}

		\begin{minipage}[t]{.49\linewidth}
			\caption{(left) Operator precedence matrix where parentheses take precedence over multiplication, which takes precedence over addition. The cells marked by $\cdot$ denote the irrelevant relations.}
			\label{table:arithmetic}
		\end{minipage}
		\hfill
		\begin{minipage}[t]{.42\linewidth}
			\caption{(center) Computation of the collapsed from of $1+0\times\lp1+1\rp$}\label{collapse:arithmetic}
			\caption{(right) Derivation tree of the words $1+0\times\lp1+1\rp \in L(G_{\text{arith}})$}\label{derive:arithmetic}
		\end{minipage}
	\end{figure}
	
	For $a,b \in \Sigma$, we write $a \popeq b$ iff $a \pop b$ or $a \shift b$, and similarly $a \pusheq b$ iff $a \push b$ or $a \shift b$.
	It is worth emphasizing that, despite their appearance, the operator precedence relations $\push, \pusheq$, $\pop$, $\popeq$ and $\shift$ are in general neither reflexive nor transitive.
	We extend the precedence relations with $\varepsilon$ such that $\varepsilon \push a$, $a \pop \varepsilon$, and $\varepsilon \shift \varepsilon$ for all $a \in \Sigma$.

	Every word induces a sequence of precedences.
	For some words, this sequence corresponds to a \emph{chain}~\cite{DBLP:journals/siamcomp/LonatiMPP15}, which is a building block of structured words.
%	Some words induce an important sequence of precedences called a \emph{chain} \cite{DBLP:journals/siamcomp/LonatiMPP15}.
	
	\begin{definition}[chain] 
		Let $a_i \in \widehat{\Sigma}$ and $u_i \in \widehat{\Sigma}^*$ for all $i \in \N$, and let $n \geq 1$.
		A word $w = a_0 a_1 \ldots a_{n+1}$ is a \emph{simple chain} when $a_0, a_{n+1} \in \widehat{\Sigma} \cup \{\varepsilon\}$ and $a_0 \push a_1 \shift a_2 \shift ... \shift a_n \pop a_{n+1}$.
		A word $w = a_0 u_0 a_1 u_1 \ldots a_n u_n a_{n+1}$ is a \emph{composite chain} when $a_0 a_1 \ldots a_{n+1}$ is a simple chain and for all $0 \leq i \leq n$, either $a_i u_i a_{i+1}$ is a (simple or composite) chain or $u_i = \varepsilon$.
		A word $w$ is a \emph{chain} when $w$ is a simple or a composite chain.
	\end{definition}
	
	For all $x, y, z \in \widehat{\Sigma}^*$, the predicate $\chain{x}{y}{z}$ holds iff $(x^{\last})y(z^{\first})$ is a chain.
	Note that, if $\chain{x}{y}{z}$ then $xyz \neq \varepsilon$.
	
	\begin{example}
		Let $\widehat{\Sigma}$ be the operator precedence alphabet in \figurename~\ref{table:arithmetic} that specifies the precedence relations for generating arithmetic expressions.
		The word $\lp \lp \rp \rp$ is a simple chain because $\lp \push \lp \shift \rp \pop \rp$.
		Moreover, the word $\lp1 + 1\rp$ is a composite chain because the words $\lp1+$, $+1\rp$, and $\lp+\rp$ are simple chains.
	\end{example}
	
	Next, we define a function that conservatively simplifies the structure of a given word.
	
	\begin{definition}[collapsing function]
		For a given operator precedence alphabet $\widehat{\Sigma}$, its \emph{collapsing function} $\lambda_{\widehat{\Sigma}} \colon \widehat{\Sigma}^* \to \widehat{\Sigma}^*$ is defined inductively as follows:
		$\lambda_{\widehat{\Sigma}}(w) = \lambda_{\widehat{\Sigma}}(xz)$ if $w = xyz$ and $\chain{x}{y}{z}$ for some $x,y,z \in \widehat{\Sigma}^+$, and $\lambda_{\widehat{\Sigma}}(w) = w$ if there is no such $x,y,z \in \widehat{\Sigma}^+$.
		When $\widehat{\Sigma}$ is clear from the context, we denote its collapsing function by $\lambda$.
	\end{definition}
	
	For every $w \in \widehat{\Sigma}$, observe that $\lambda(w)$ is in the following collapsed form: there exist $1 \leq i \leq j \leq n = |\lambda(w)|$ such that $a_1 \popeq \ldots \popeq a_{i-1} \pop a_i \shift a_{i+1} \shift \ldots \shift a_j \push a_{j+1} \pusheq \ldots \pusheq a_n$.
	
	\begin{example}
		Let $\widehat{\Sigma}$ be the operator precedence alphabet in \figurename~\ref{table:arithmetic}.
		Let $w = \lp1 + 0\rp \times \lp1 + 1\rp$ and observe that $\lambda(w) = \lp\rp \times \lp\rp$ since $\chain{\lp}{1+0}{\rp}$ and $\chain{\lp}{1+1}{\rp}$.
		Note also that $\lp \shift \rp \pop \times \push \lp \shift \rp$.
	\end{example}

	Note that the collapsed form is unique and allows us to generalize classical notions of well-nested words.
	
	\begin{definition} [structured words]
		Let $\widehat{\Sigma}$ be an operator precedence alphabet.
		We define the following sets of words:
		\[\begin{array}{ll}
			\widehat{\Sigma}^*_{\pusheq} &= \{w \in \widehat{\Sigma}^* ~|~ \lambda(w) = a_1 \ldots a_n \text{ where $ a_{i} \pusheq a_{i+1}$ for all $i$, or } |\lambda(w)| \leq 1 \}
			\\
			\widehat{\Sigma}^*_{\popeq} &=  \{w \in \widehat{\Sigma}^* ~|~ \lambda(w) = a_1\ldots a_n \text{ where $ a_{i} \popeq a_{i+1}$ for all $i$, or } |\lambda(w)| \leq 1 \}
			\\
			\widehat{\Sigma}^*_{\shift} &=  \{w \in \widehat{\Sigma}^* \st \lambda(w) = a_1 \ldots a_n \text{ where $ a_{i} \shift a_{i+1}$ for all $i$, or } |\lambda(w)| \leq 1 \} = \widehat{\Sigma}^*_{\pusheq} \cap \widehat{\Sigma}^*_{\popeq}
		\end{array}\]
		%that we respectively call \emph{right-matched}, \emph{left-matched}, and \emph{well-matched} words over $\widehat{\Sigma}$.
	\end{definition}

	Looking back at the definition of collapsed form, one can verify for every word $w \in \widehat{\Sigma}^*$ that $w \in \widehat{\Sigma}^*_{\pusheq}$ iff $i = 1$, and $w \in \widehat{\Sigma}^*_{\popeq}$ iff $j = n$.
	
	\begin{example}
		Let $\widehat{\Sigma}$ be the operator precedence alphabet in \figurename~\ref{table:arithmetic}.
		The word $+ \times \lp \rp$ is in $\widehat{\Sigma}^*_{\pusheq}$, the word $\lp\rp \times +$ is in $\widehat{\Sigma}^*_{\popeq}$, and the word $\lp\rp$ is in $\widehat{\Sigma}^*_{\shift}$.
		Moreover, note that $+ \push \times \push \lp \shift \rp$ and $\lp \shift \rp \pop \times \pop +$.
	\end{example}

	\subsection{Operator Precedence Grammars}
	
	A \emph{context-free grammar} $G = (\Sigma, V, R, S)$ is tuple where $\Sigma$ is a finite set of terminal symbols, $V$ is a finite set of non-terminal symbols, $R \subseteq V \times (\Sigma \cup V)^*$ is a finite set of derivation rules, and $S \in V$ is the starting symbol.
	Given $\alpha, \beta \in (\Sigma \cup V)^*$, we write $\alpha \rightarrow \beta$ when $\beta$ can be derived from $\alpha$ with one rule, i.e., when there exists $(\alpha_2, \beta_2) \in R$, $\alpha = \alpha_1\alpha_2\alpha_3$ and $\beta=\alpha_1\beta_2\alpha_3$.
	Derivations using a sequence of rules are denoted by $\rightarrow^*$, the transitive closure of the relation $\rightarrow$.
	The language of $G$ is $L(G) = \{ w \in \Sigma^* \st S \rightarrow^* w\}$.
	A derivation tree for $u \in L(G)$ is a tree over $\Sigma \cup V \cup \{\varepsilon\}$ such that the root is labeled by $S$, the concatenation of all leaves is $u$, and if a node is labeled by $\alpha$ and its children labeled by $\beta_1, \dots, \beta_k$ then $(\alpha, \beta_1 \dots \beta_k) \in R$.
	A grammar is said to be \emph{non-ambiguous} when for all $u \in L(G)$ admits a unique derivation tree.
	
	Intuitively, an \emph{operator precedence grammar} (OPG for short) is an unambiguous context-free grammar whose derivation trees comply with some operator precedence matrix.
	Formally, let $G = (\Sigma, V, R, S)$ be a context-free grammar and $A \in V$ be a non-terminal, and define the following sets of terminal symbols where $B \in V \cup \{\varepsilon\}$ and $\alpha \in (V \cup \Sigma)^*$:
		\[\mathcal{L}_G(A) = \{a \in \Sigma \st A \rightarrow^* B a \alpha\} \hspace*{1cm} \mathcal{R}_G(A) = \{a \in \Sigma \st A \rightarrow^* \alpha a B\}\]
	Given $a,b \in \Sigma$, we define the following operator precedence relations where $\alpha,\beta \in (V \cup \Sigma)^*$:
	\begin{itemize}
		\item $a \push_G b$ iff there exists a rule $A \to \alpha a C \beta$ where $C \in V$ and $b \in \mathcal{L}_G(C)$,
		\item $a \pop_G b$ iff there exists a rule $A \to \alpha C b \beta$ where $C \in V$ and $a \in \mathcal{R}_G(C)$,
		\item $a \shift_G b$ iff there exists a rule $A \to \alpha a C b \beta$ where $C \in V \cup \{\varepsilon\}$.
	\end{itemize}
	Finally, $G$ is an operator precedence grammar if and only if for all $a,b \in \Sigma$, we have that $|\{ \odot \in \{\push_G, \shift_G, \pop_G\} \st a \odot b\}| \leq 1$.

	\begin{example} \label{ex:prelim:arithmetic}
		Let $G_{\text{arith}} = (\Sigma, V, R, A)$ be a context-free grammar over $\widehat{\Sigma} = \{+, \times, \lp, \rp, 0, 1 \}$ as in \figurename~\ref{table:arithmetic} where $V = \{A, B, C\}$ and $R$ contains the following rules:
		\[A \to A + B ~|~ B \hspace*{1cm} B \to B \times C ~|~ C \hspace*{1cm}  C \to \lp A \rp ~|~ 0 ~|~ 1\]
		The language $L(G_{\text{arith}})$ consists of valid arithmetic expressions with an implicit relation between terminal symbols: parentheses take precedence over multiplication, which takes precedence over addition~\cite{DBLP:journals/siamcomp/LonatiMPP15}.
		The missing relations, replaced by $\cdot$ in the matrix of \figurename~\ref{table:arithmetic}, denote the precedence relations that cannot be encountered by the given grammar, so the chosen precedence relation does not matter.
		For example, $00$ and $\rp\lp$ are not valid arithmetic expressions and cannot be generated by $G_{\text{arith}}$.
		We remark that the structures of derivation trees and chains share strong similarities as highlighted by \figurename~\ref{collapse:arithmetic} and \figurename~\ref{derive:arithmetic}.
	\end{example}

	\subsection{Operator Precedence Automata}
	
	Intuitively, operator precedence automata are pushdown automata where stack operations are determined by the precedence relations between the next letter and the top of the stack.
	
	\begin{definition} [operator precedence automaton]	
		An \emph{operator precedence automaton} (OPA for short) over $\widehat{\Sigma}$ is a tuple $\mathcal{A} = (Q, I, F, \Delta)$ where $Q$ is a finite set of states, $I \subseteq Q$ is the set of initial states, $F \subseteq Q$ is a set of accepting states, and	$\Delta \subseteq \big(Q \times (\Sigma \cup \{\varepsilon\}) \times (\Gamma^+ \cup \{\bot\})\big)^2$ is the $\widehat{\Sigma}$-driven transition relation where $\Gamma = \Sigma \times Q$ is the stack alphabet and $\bot$ denotes the empty stack, meaning that, when $((s, a, \alpha), (t, b, \beta)) \in \Delta$ the following holds:
		\begin{itemize}
			\item If $\alpha = \bot$ or $\alpha=\langle q,a' \rangle\alpha'$ with $a' \push a$, then the input triggers a \emph{push} stack-operation implying that $b=\varepsilon$ and $\beta = \langle s,a \rangle\alpha$.
			We write $(s,\alpha) \tikzpush{a} (t, \beta)$.
			\item If $\alpha=\langle q,a' \rangle\alpha'$ with $a' \shift a$, then the input triggers a \emph{shift} stack-operation implying that $b=\varepsilon$ and $\beta = \langle q,a \rangle\alpha'$.
			We write $(s,\alpha) \tikzshift{a} (t, \beta)$.
			\item If $\alpha=\langle q,a' \rangle\alpha'$ with $a' \pop a$, then the input triggers a \emph{pop} stack-operation implying that $b=a$ and $\beta = \alpha'$.
			We write $(s,\alpha) \tikzpop{a} (t, \beta)$.
		\end{itemize}
	\end{definition}
	
	Let $\A$ be an OPA.
	A \emph{configuration} of $\A$ is a triplet $(q, u, \theta)$ where $q \in Q$ is the current state, $u\in\Sigma^*$ is the input suffix left to be read, and $\theta \in \Gamma^+\cup\{\bot\}$ is the current stack.
	A \emph{run} of $\A$ is a finite sequence of configurations $\left((q_{i}, u_{i}, \theta_{i})\right)_{1 \leq i \leq n}$ for some $n \in \N$ such that, for all $1 \leq i \leq n$, the automaton fires (i) a push-transition $(q_{i-1}, \theta_{i-1}) \tikzpush{a} (q_i, \theta_i)$ where $u_{i-1}=au_i$, (ii) a shift-transition $(q_{i-1}, \theta_{i-1}) \tikzshift{a} (q_i, \theta_i)$ where $u_{i-1}=au_i$, or (iii) a pop-transition $(q_{i-1}, \theta_{i-1}) \tikzpop{a} (q_i, \theta_i)$ where $u_{i-1}=u_i \in \{au \st u\in\Sigma^*\}$.
	We write $(s, u, \alpha)\tikztransition (t, v, \beta)$ when $(s, u, \alpha)(t, v, \beta)$ is a run, and let $(s, u, \alpha) \tikzpath (t, v, \beta)$ be its reflexive transitive closure.
	For all $n \in\N$, we define the predicate $(s, u, \alpha) \tikzpath[n] (t, v, \beta)$ inductively %\rednote{Haven't found a place where this is used. Is it necessary? - PK} it's used in the cat function for the antichain algorithm
	by $(s, u, \alpha) = (t, v, \beta)$ when $n=0$ and
	by  $\exists (q, w, \theta), (s, u, \alpha) \tikzpath[] (q, w, \theta) \tikzpath[n-1] (t, v, \beta)$ otherwise. 
	The \emph{language} of $\A$ is defined by $ L(\A) = \{w \in \Sigma^* \st q_0\in I, q_F\in F, (q_0, w, \bot) \tikzpath (q_F, \varepsilon, \bot)\}$.
	An OPA is \emph{deterministic} when $|I|=1$ and $\Delta$ is a function from $Q \times \Sigma \times (\Gamma^+ \cup \{\bot\})$ to $Q\times (\Sigma \cup \{\varepsilon\}) \times (\Gamma^+ \cup \{\bot\})$, %\rednote{I'm not sure whether this is correct. What do you mean by ``to itself''? Is it something different than the default definition of $\Delta$? If the default definition of $\Delta$ should be non-deterministic, then it should be in a different form. - PK} when the automaton is deterministic, the transition relation becomes a function whose domain equals its range
	and it is \emph{complete} when from every configuration $(s, u, \theta)$ there exists a run that ends in $(t, \varepsilon, \bot)$ for some state $t \in Q$.
	For a given stack $\theta \in \Gamma^+ \cup \{\bot\}$, we define $\theta^\top$ as the stack symbol at the top of $\theta$ if $\theta \in \Gamma^+$, and $\theta^\top = \varepsilon$ if $\theta=\bot$.
	
	\begin{definition} [operator precedence language]
		An \emph{operator precedence language} (OPL for short) is a language recognized by some operator precedence automaton.
	\end{definition}
	
	If $L$ is an OPL over the operator precedence alphabet $\widehat{\Sigma}$, we say that $L$ is a $\widehat{\Sigma}$-OPL.
	
	\begin{remark}
		The literature on OPLs often assumes the $\shift$-acyclicity of operator precedence relations of the alphabet, i.e., that there is no $n \geq 1$ and $a_1,\ldots,a_n \in \Sigma$ with $a_1 \shift \ldots \shift a_n \shift a_1$.
		This assumption is used to bound the right-hand side of OPG derivation rules, and find a key application for constructing an OPG that recognizes the language of a given OPA~\cite{DBLP:journals/siamcomp/LonatiMPP15}.
		We omit this assumption since it is not needed for establishing the results on OPAs, including the construction of an OPA that recognizes the language of a given OPG.
	\end{remark}
	
	Now, we present an OPA that recognizes valid arithmetic expressions.
	
\begin{figure}[t]\centering
		\noindent\begin{minipage}[c]{.25\linewidth}\centering
			\includegraphics[scale=0.5]{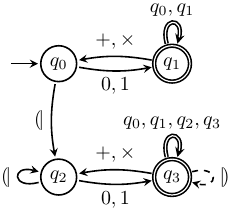}
%			\scalebox{0.75}{
%				\begin{tikzpicture}[node distance =2cm]
%					\node[state, initial, initial angle=180, label=center:$q_0$] (0) {};
%					\node[state, double, right of = 0, label=center:$q_1$] (1) {};
%					\node[state, below of = 0, label=center:$q_2$] (2) {};
%					\node[state, double, right of = 2, label=center:$q_3$] (3) {};
%					
%					\path[transition]
%					(0) edge[bend right=10] node[below] {$0, 1$} (1)
%					(1) edge[bend right=10] node[above] {$+,\times$} (0)
%					(1) edge[loop above, double] node[above] {$q_0,q_1$} (1)
%					(0) edge[bend right=10] node[left] {$\lp$} (2)
%					(2) edge[loop left] node[left] {$\lp$} (2)
%					(2) edge[bend right=10] node[below] {$0, 1$} (3)
%					(3) edge[bend right=10] node[above] {$+,\times$} (2)
%					(3) edge[loop above, double] node[above] {$q_0,q_1,q_2,q_3$} (3)
%					(3) edge[loop right, dashed] node[right] {$\rp$} (3)
%					;	
%				\end{tikzpicture}
%			}
		\end{minipage}
		\begin{minipage}[c]{.74\linewidth}\centering
%			\scalebox{0.75}{
%				\begin{tabular}{c|r|r}
%					state & current input & \multicolumn{1}{c}{stack} \\ \hline
%					$q_0$ & $1 \times (0 + 1) $ & $\bot$ \\ \hline
%					$q_1$ & $\times (0 + 1) $ & $\stack{q_0,1} \bot$ \\ \hline
%					$q_1$ & $\times (0 + 1) $ & $\bot$ \\ \hline
%					$q_0$ & $(0 + 1) $ & $\stack{q_1,{\times}} \bot$ \\ \hline
%					$q_2$ & $0 + 1) $ & $\stack{q_0,{(}} \stack{q_1,{\times}} \bot$ \\ \hline 
%					$q_3$ & $+ 1) $ & $\stack{q_2,0} \stack{q_0,{(}} \stack{q_1,{\times}} \bot$ \\ \hline 
%					$q_3$ & $+ 1) $ & $\stack{q_0,{(}} \stack{q_1,{\times}} \bot$ \\ \hline 
%					$q_2$ & $1) $ & $\stack{q_3,{+}} \stack{q_0,{(}} \stack{q_1,{\times}} \bot$ \\ \hline 
%					$q_3$ & $) $ & $\stack{q_2,1} \stack{q_3,{+}} \stack{q_0,{(}} \stack{q_1,{\times}} \bot$ \\ \hline
%					$q_3$ & $) $ & $\stack{q_3,{+}} \stack{q_0,{(}} \stack{q_1,{\times}} \bot$ \\ \hline
%					$q_3$ & $) $ & $\stack{q_0,{(}} \stack{q_1,{\times}} \bot$ \\ \hline
%					$q_3$ & $\varepsilon $ & $\stack{q_0,)} \stack{q_1,{\times}} \bot$ \\ \hline
%					$q_3$ & $\varepsilon $ & $\stack{q_1,{\times}} \bot$ \\ \hline
%					$q_3$ & $\varepsilon $ & $\bot$ \\
%			\end{tabular}}
		\scalebox{0.75}{
			\begin{tabular}{c|r|r||c|r|r}
				state & \multicolumn{1}{c|}{input} & \multicolumn{1}{c||}{stack} & state & \multicolumn{1}{c|}{input} & \multicolumn{1}{c}{stack} \\ \hline
				$q_0$ & $1 \times \lp0 + 1\rp $ & $\bot$ & $q_2$ & $1\rp$ & $\stack{q_3,{+}} \stack{q_0,{\lp}} \stack{q_1,{\times}} \bot$ \\ \hline 
				$q_1$ & $\times \lp0 + 1\rp$ & $\stack{q_0,1} \bot$ & $q_3$ & $\rp $ & $\stack{q_2,1} \stack{q_3,{+}} \stack{q_0,{\lp}} \stack{q_1,{\times}} \bot$ \\ \hline
				$q_1$ & $\times \lp0 + 1\rp $ & $\bot$ & $q_3$ & $\rp $ & $\stack{q_3,{+}} \stack{q_0,{\lp}} \stack{q_1,{\times}} \bot$ \\ \hline
				$q_0$ & $\lp0 + 1\rp $ & $\stack{q_1,{\times}} \bot$ &	$q_3$ & $\rp$ & $\stack{q_0,{\lp}} \stack{q_1,{\times}} \bot$ \\ \hline
				$q_2$ & $0 + 1\rp $ & $\stack{q_0,{\lp}} \stack{q_1,{\times}} \bot$ & $q_3$ & $\varepsilon $ & $\stack{q_0,\rp} \stack{q_1,{\times}} \bot$ \\ \hline
				$q_3$ & $+ 1\rp $ & $\stack{q_2,0} \stack{q_0,{\lp}} \stack{q_1,{\times}} \bot$ & $q_3$ & $\varepsilon $ & $\stack{q_1,{\times}} \bot$ \\ \hline
				$q_3$ & $+ 1\rp $ & $\stack{q_0,{\lp}} \stack{q_1,{\times}} \bot$ & $q_3$ & $\varepsilon $ & $\bot$
		\end{tabular}}
		\end{minipage}
		\caption{An OPA recognizing the arithmetic expressions generated by the OPG in Example~\ref{ex:prelim:arithmetic} and its run on the input word $1 \times \lp0 + 1\rp$. Shift-, push-, and pop-transitions are respectively denoted by dashed, normal, and double arrows.} \label{fig:prelim:arithmetic}
	\end{figure}

	\begin{example}
		Recall the OPG of Example~\ref{ex:prelim:arithmetic} generating arithmetic expressions over the operator precedence alphabet of \figurename~\ref{table:arithmetic}.
		In \figurename~\ref{fig:prelim:arithmetic}, we show an OPA that recognizes the same language and an example of a computation.
	\end{example}
	
	\subsection{Expressiveness and Decidability of Operator Precedence Languages}

	In this section, briefly summarize some known results about OPLs.
	First, we remark that OPLs are context-free languages as they are recognized by a subclass of pushdown automata.
		
	\begin{theorem}[from~\cite{DBLP:journals/jcss/Crespi-ReghizziM12}]
		Deterministic context-free languages strictly include OPLs.
	\end{theorem}
	
	The language $L = \{a^n b a^n \st n \geq 0\}$, which is a deterministic context-free language, separates the two classes.
	Indeed, it is not an OPL because while the first segment of $a^n$ must push to the stack (i.e., $a \push a$), the last segment must pop (i.e., $a \pop a$), resulting in conflicting precedence relations.
	Next, we recall that OPLs enjoy the many closure properties.
	
	\begin{theorem}[from~\cite{DBLP:journals/jcss/Crespi-ReghizziM12,DBLP:journals/iandc/Crespi-ReghizziMM78}]
		OPLs are closed under boolean operations, concatenation, Kleene star, reversal, prefixing, and suffixing.
	\end{theorem}

	The class of VPLs enjoy these closure as well.
	In fact, every VPL can be expressed as an OPL with an operator precedence alphabet designed as follows: internal characters and returns take precedence over any character; calls equal in precedence with returns, and they yield precedence to calls and internal characters.
	
	\begin{theorem}[from~\cite{DBLP:journals/jcss/Crespi-ReghizziM12}]
		OPLs strictly include visibly pushdown languages.
	\end{theorem}

	The language $L = \{a^n b^n \st n \geq 1\} \cup \{c^n d^n \st n \geq 1\} \cup \{e^n(bd)^n \st n \geq 1\}$, which is an OPL due to their closure under union, separate the two classes.
	Indeed, for $L$ to be a VPL, the first set requires that $a$ is a call and $b$ is a return.
	Similarly, $c$ is a call and $d$ is a return due to the second set.
	However, the last set requires that at most one
	of $b$ and $d$ is a return, resulting in a contradiction.
	We also note that OPAs support determinization.
	
	\begin{theorem}[from~\cite{DBLP:journals/siamcomp/LonatiMPP15}]
		Every OPL can be recognized by a deterministic OPA.
	\end{theorem}

	Despite their expressive power, OPL remain decidable for the classical decision problems.
	In particular, OPAs enjoy the same order of complexity as VPDA for basic decision problems.
	
	\begin{theorem}[from~\cite{DBLP:journals/ipl/Lange11,DBLP:journals/siamcomp/LonatiMPP15}]\label{thm:OPAcomplexity}
		The language emptiness is in \textsc{PTime-C} for OPAs.
		The language inclusion, universality, and equivalence are in \textsc{PTime} for deterministic OPAs and \textsc{ExpTime-C} for nondeterministic OPAs.
	\end{theorem}

	\begin{remark} \label{rem:OPAcomplexity}
		The membership problem is in \textsc{PTime} for OPAs.
		Determining whether a given word $w$ is accepted by a given OPA $\A$ can be done in polynomial time by constructing an automaton $\B$ that accepts only $w$, constructing the intersection $\C$ of $\A$ and $\B$, and deciding the non-emptiness of $\C$.
	\end{remark}

	\section{A Finite Congruence for Operator Precedence Languages}

This section introduces a congruence-based characterization of OPLs, similar to the Myhill-Nerode congruence for regular languages.
We let $\widehat{\Sigma}$ be an operator precedence alphabet throughout the section.
A relation $\bow$ over $\widehat{\Sigma}^*$ is monotonic when $x \bow y$ implies $uxv \bow uyv$ for all $x, y, u, v \in \widehat{\Sigma}^*$.
Intuitively, monotonicity requires two words in relation to stay related while becoming embedded into some context that constructs a larger word.
However, such a definition is not well suited for structured  words as it does not follow how chains are constructed.
Hence, we introduce a more restrictive notion than monotonicity.

\begin{definition}[chain-monotonicity]
	A relation $\bow$ over $\widehat{\Sigma}^*$ is \emph{chain-monotonic} when $x~\bow~y$ implies $uu_0xv_0v \bow uu_0yv_0v$ for all $x, y, u, v, u_0, v_0 \in \widehat{\Sigma}^*$ such that $u_0z^{\first}\in \widehat{\Sigma}^*_{\popeq}$, $z^{\last}v_0\in \widehat{\Sigma}^*_{\pusheq}$, and $\chain{u}{u_0zv_0}{v}$ for each $z\in\{x, y\}$.
\end{definition}

Chain-monotonicity requires
%monotonicity for embeddings into larger words that are ``structured'' in a specific way. 
two words in relation to stay related while being embedded into some context that construct larger structured words.
This leads us to describe when two words agree on whether an embedding into a larger word forms a chain.
For this, we introduce a relation that relates words that behave similarly with respect to the chain structure.

\begin{definition}[chain equivalence]
	We define the \emph{chain equivalence} $\approx$ over $\widehat{\Sigma}^*$ as follows:
	\[
		x \approx y \iff
		 \bigwedge \begin{cases}
		 	x^{\first} = y^{\first} \land x^{\last} = y^{\last}\\
			\forall u, v, u_0, v_0 \in \widehat{\Sigma}^*, \big(u_0x^{\first}\in \widehat{\Sigma}^*_{\popeq} \land x^{\last}v_0\in \widehat{\Sigma}^*_{\pusheq}\big) \Rightarrow \big(\chain{u}{u_0xv_0}{v} \Leftrightarrow \chain{u}{u_0yv_0}{v}\big)
		\end{cases}
	\]
\end{definition}

%We note that the conditions $x^{\first} = y^{\first}$ and $x^{\last} = y^{\last}$  are too fine for the purpose of the chain equivalence.
%One can make the chain equivalence coarser by relaxing these conditions while preserving transitivity.
%However, for technical convenience, we keep these conditions instead.
We observe that $\varepsilon$ is in relation with itself exclusively, i.e., $x=\varepsilon$ iff $\varepsilon \approx x$ iff $x \approx \varepsilon$. %\rednote{Is the last part necessary? ($x \approx \varepsilon$) Symmetry is mentioned below. - PK} doesn't hurt to repeat in math symbols, people often skim these texts :)
Consider a word $w \in \widehat{\Sigma}^+$ for which $\lambda(w)$ is of the form $a_1\dots a_\ell b_1 \dots b_m c_1 \dots c_n$ for some $\ell, m, n\in \N$ such that
$a_1 \popeq \ldots \popeq a_\ell \pop b_1 \shift \ldots \shift b_m \push c_1 \pusheq \ldots \pusheq c_n$
where $a_i, b_j, c_k \in \Sigma$ for all $i, j, k$.
We define the \emph{profile} of $w$ as $P_w = (w^{\first}, w^{\last}, P_w^{\first}, P_w^{\last})$, where
$P_w^{\first} = \{a_1, b_1\} \cup \{a_{i+1} \st a_i \pop a_{i+1}, 1 \leq i < \ell\}$ and
$P_w^{\last} = \{b_m, c_n\} \cup \{c_k \st c_k \push c_{k+1}, 1 \leq k < n\}$.
There are at most $|\Sigma|^2\times 2^{2|\Sigma| - 2} + 1$ profiles.
We can show that two words with the same profile are chain equivalent, leading to the following proposition.

\begin{proposition}\label{proposition:chain:finite}
	$\approx$ is a chain-monotonic equivalence relation with finitely many classes.
\end{proposition}

Next, we introduce an equivalence relation that characterizes OPLs.
	
	\begin{definition}[syntactic congruence]
		Given $L\subseteq \widehat{\Sigma}^*$, we define $\equiv_L$ as the following relation over $\widehat{\Sigma}^*$:
		\[x \equiv_L y \iff x \approx y \land 
		\begin{cases}
			\forall u, v, u_0, v_0 \in \widehat{\Sigma}^*,
			\big(u_0x^{\first}\in \widehat{\Sigma}^*_{\popeq} \land x^{\last}v_0\in \widehat{\Sigma}^*_{\pusheq} \land \chain{u}{u_0xv_0}{v}\big)\\
			\qquad \Rightarrow \big(uu_0xv_0v\in L \Leftrightarrow uu_0yv_0v\in L\big)
		\end{cases}
		\]
	\end{definition}

	Let us demonstrate the syntactic congruence.

	\begin{example}
		Let $\Sigma = \{a,b\}$ and let $\widehat{\Sigma}$ be the operator precedence alphabet with the relations $a \push a$, $a \shift b$, $b \pop a$, and $b \pop b$.
		Consider the language $L = \{a^n b^n \st n \geq 1\}$.
		
		There are 17 potential profiles for $\widehat{\Sigma}$ in total.
		Although some of them cannot occur due to the precedence relations of $\widehat{\Sigma}$, the remaining ones correspond to the equivalence classes of ${\approx}$.
		For example, $(a, a, \{a\}, \{a,b\})$ cannot occur since $b \pop a$, and $(a, b, \{a\}, \{b\})$ contains exactly the words in $L$ which are of the form $a^n b^n$ for some $n \geq 1$.
		For brevity, we only show how the syntactic congruence ${\equiv_L}$ refines the class of ${\approx}$ corresponding to $(a, a, \{a\}, \{a\})$ by splitting it into four subclasses.
		The profile $(a, a, \{a\}, \{a\})$ captures exactly the words of the form $w = a$ or $w = a u a$ where in each prefix of $au$ there are no more $b$'s than $a$'s.
		Notice that for such $w$, $\lambda(w)$ is of the form $(ab)^*a^+$, where $a^+=\{a^n\mid n>0\}$.
	
		We first argue that $a \not \equiv_L aa$ but $aa \equiv_L aa^n$ for all $n \geq 1$.
		Taking $u = v = u_0 = \varepsilon$ and $v_0 = b$, observe that the preconditions for the syntactic congruence are satisfied but $ab \in L$ while $aab \notin L$, therefore $a \not \equiv_L aa$.
		Now, let $n \geq 2$, and consider the words $aa$ and $aa^n$.
		Intuitively, since there is no $x,y \in \widehat{\Sigma}^*$ such that $x aa y \in L$ and $x aa^n y \in L$, we show that whenever the preconditions for the congruence are satisfied, both longer words are out of $L$.
		Given $u,v,u_0,v_0 \in \widehat{\Sigma}^*$ such that $u_0 a \in \widehat{\Sigma}^*_{\popeq}$, $a v_0 \in \widehat{\Sigma}^*_{\pusheq}$, and $\chain{u}{u_0 aa v_0}{v}$, we assume towards contradiction that $u u_0 aa v_0 v \in L$.
		%Then, there exists $k \in \N$ such that $u u_0 = a^k$ and $v_0 v = b^{k+2}$.
		Since $u u_0 aa v_0 v \in L$ and $u_0 a \in \widehat{\Sigma}^*_{\popeq}$, we have $u_0=\varepsilon$.
		Moreover, since $a v_0 \in \widehat{\Sigma}^*_{\pusheq}$, we have that $v_0$ is either of the from $a^*$ or $a^*b$.
		Consequently, $\lambda(u_0aav_0)$ is $aaa^*$ or $aaa^*b$.
		This contradicts that $\chain{u}{u_0 aa v_0}{v}$ because $a \push a$, and therefore $u u_0 aa v_0 v \notin L$.
		The same argument shows that $u u_0 aa^n v_0 v \notin L$, implying that $aa \equiv_L aa^n$.
		Similarly as above, we can show that $u \not \equiv_L v$ but $v \equiv_L w$ for all $u, v, w \in \widehat{\Sigma}^*$ such that $\lambda(u) = (ab)^ia$, $\lambda(v) = (ab)^jaa$, and $\lambda(w) = (ab)^kaa^n$, where $n, i, j, k \geq 1$.
%		Now, let $w_1,w_2 \in \widehat{\Sigma}^*$ be of the form $aba^+$ such that $\lambda(w_1) = aba$ and $\lambda(w_2) = abaa^k$ for some $k \geq 1$.
%		Since for every choice of $u,v,u_0,v_0 \in \widehat{\Sigma}^*$ we have $u u_0 w_1 v_0 v \notin L$ and $u u_0 w_2 v_0 v \notin L$, we get $w_1 \equiv_L w_2$.	
%		Finally, simply observe that the classes $[u]$ and $[v]$ are different from $[a]$ and $[aa]$.
%		There are 17 profiles for $\widehat{\Sigma}$ in total.
%		For every $w \in \widehat{\Sigma}^*$, we argue that
%		(i) if $w^{\last} = a$ then $b \notin P_w^{\last}$,
%		(ii) if $w^{\first} = b$, $w^{\last} = b$, and $a \in P_w^{\last}$ then $a \in P_w^{\first}$, and
%		(iii) if $w^{\first} = b$ and $w^{\last} = a$ then $a \in P_w^{\first}$.
%		Intuitively, (i) is because $b \pop a$.
%		Moreover, (ii) and (iii) are because $b_1 = b_m = a$ for such $w$.
%		Then, the equivalence relation $\approx$ yields 11 classes, each of which corresponding to one of the following profiles:
%		$(\varepsilon, \varepsilon, \{\varepsilon\}, \{\varepsilon\})$,
%		$(a, a, \{a\}, \{a\})$,
%		$(a, a, \{a,b\}, \{a\})$,
%		$(b, b, \{b\}, \{b\})$,
%		$(b, b, \{a,b\}, \{b\})$,
%		$(b, b, \{a,b\}, \{a,b\})$,
%		$(b, a, \{b,a\}, \{a\})$,
%		$(a, b, \{a\}, \{b\})$,
%		$(a, b, \{a\}, \{a,b\})$,
%		$(a, b, \{a,b\}, \{a\})$,
%		$(a, b, \{a,b\}, \{a,b\})$.
	\end{example}

	We now show that the syntactic congruence is chain-monotonic.

	\begin{theorem} \label{thm:congruencechainmono}
		For every $L\subseteq \widehat{\Sigma}^*$, $\equiv_L$ is a chain-monotonic equivalence relation.
	\end{theorem}

	The main result of this section is the characterization theorem below.
	We prove each direction separately in Sections 3.1 and 3.2.
	
	\begin{theorem}\label{main:syntactic}
		A language $L$ is an OPL iff ${\equiv_L}$ admits finitely many equivalence classes.
	\end{theorem}	
	
	\subsection{Finiteness of the Syntactic Congruence}
	
	%TODO: check
	Let $\widehat{\Sigma}$ be an operator precedence alphabet, $\A = (Q, I, F, \Delta)$ be an OPA over $\widehat{\Sigma}$, and $\star \notin \Sigma$ be a fresh letter for which we extend the precedence relation with $a \push \star$ for all $a \in \Sigma$.
	
	For every word $w\in \widehat{\Sigma}^*$, we define the functions $f_w \colon Q \times (\Gamma\cup\{\bot\}) \rightarrow 2^Q$ %\rednote{Why not put the range to be $Q$? For deterministic OPA, it will be singleton anyway and furthermore, the following proof looks as it assumes it is a singleton. The same applies for $\varPhi$ and $g$. However, one would then need to restrict the set of words from $\widehat{\Sigma}$ for which the function is defined. - PK.} we need the nondeterminism for the antichain section and this just makes it easier to reuse the definitions
	and $\varPhi_w \colon Q \times (\Gamma\cup\{\bot\}) \rightarrow 2^{\Gamma^+ \cup \{\bot\}}$ such that for all $q \in Q$ and all $\gamma \in \Gamma\cup\{\bot\}$, we have $f_w(q, \gamma) = \{q_w \in Q \st \exists \gamma_w \in \Gamma^+\cup\{\bot\}, (q, w\star, \gamma) \tikzpath (q_w, \star,\gamma_w) \}$ and $\varPhi_w(q, \gamma) = \{\gamma_w \in \Gamma^+\cup\{\bot\} \st \exists q_w \in Q, (q, w\star, \gamma) \tikzpath (q_w, \star,\gamma_w) \}$.
	%Note that the stacks in $\varPhi_w(q, \gamma)$ may modify the given bottom $\gamma$.
	Intuitively, the states in $f_w(q, \gamma)$ and the stacks in $\varPhi_w(q, \gamma)$ come from the configurations that $\A$ can reach after reading $w$ from the state in $q$, but before triggering any pop-transition due to reaching the end of the word $w$.
	
	Furthermore, for every $w\in \widehat{\Sigma}^*$, we define the function $g_w \colon Q^2 \times (\Gamma\cup\{\bot\}) \rightarrow 2^Q$ such that for all $q_1, q_2 \in Q$ and all $\gamma \in \Gamma\cup\{\bot\}$ we have $g_w(q_1, q_2, \gamma) = \{ p_w \in Q \st \exists \gamma_w \in \varPhi_w(q_1, \gamma), (q_2, \varepsilon, \gamma_w) \tikzpath (p_w, \varepsilon, \bot) \}$.
	Intuitively, $g_w(q_1, q_2, \gamma)$ is the set of states that $\A$ can reach after triggering from $q_2$ the pop-transitions that empty the (unique) stack $\gamma_w \in \varPhi_w(q_1, \gamma)$ that was generated by reading $w$ while moving from the state $q_1$ to some state in $f_w(q_1, \gamma)$.
	
	Recall that for a given stack $\theta \in \Gamma^+ \cup \{\bot\}$, we denote by $\theta^\top$ the stack symbol at the top of $\theta$, which is $\varepsilon$ when $\theta=\bot$.
	Moreover, for a given set of stacks $\Theta \subseteq \Gamma^+ \cup \{\bot\}$, let us define $\Theta^\top = \{\theta^\top \st \theta \in \Theta\}$.
	For the sequel, we define the following equivalence relation:
	
	\begin{definition}[structural congruence]
		Given an OPA $A=(Q, I, F, \Delta)$, we define the relation $\equiv_A$ over $\widehat{\Sigma}^*$ as follows:
		\[x \equiv_{\A} y \iff x \approx y \land f_x = f_y \land g_x = g_y \land \big( \forall q \in Q, \forall \gamma \in \Gamma \cup \{\bot\}, (\varPhi_{x}(q,\gamma))^\top = (\varPhi_{y}(q,\gamma))^\top \big)\]
	\end{definition}
	
	First, we show that the structural congruence of any OPA has a finite index.
	
	\begin{lemma} \label{lem:finite}
		For every OPA $\A$ with $n$ states and $m$ input letters, the structural congruence $\equiv_{\A}$ has at most 
		$\mathcal{O}(m)^{\mathcal{O}(m \times n)^{\mathcal{O}(1)}}$ equivalence classes.
	\end{lemma}

	Then, we show that for any OPA the syntactic congruence of its language is coarser than its structural congruence, therefore has a finite index as well.
		
	\begin{lemma} \label{lem:coarse}
		For every OPA $\A$, the congruence $\equiv_{L(\A)}$ is coarser than the congruence $\equiv_{\A}$.
	\end{lemma}

	As a direct result of Lemmas~\ref{lem:finite}~and~\ref{lem:coarse} above, we obtain the following.
	
	\begin{corollary}
		For every $L \subseteq \widehat{\Sigma}^*$, if $L$ is a $\widehat{\Sigma}$-OPL then $\equiv_L$ has finite index.
	\end{corollary}

	\subsection{From the Syntactic Congruence to Operator Precedence Automata}
	
	%Now, we prove the other direction to characterize OPLs by the finiteness of their syntactic congruence.
	
	Consider a language $L \subseteq \widehat{\Sigma}^*$ such that $\equiv_L$ has finitely many equivalence classes.
	We construct a deterministic OPA that recognizes $L$ and whose states are based on the equivalence classes of ${\equiv_L}$.
	Given $w\in\widehat{\Sigma}^*$, we denote by $[w]$ its equivalence class with respect to ${\equiv_L}$.
	We construct $\A = (Q, \{q_0\}, F, \Delta)$ with the set of states $Q = \{([u], [v]) \st u, v\in \widehat{\Sigma}^*\}$, the initial state $q_0 = ([\varepsilon], [\varepsilon])$, the set of accepting states $F = \{([\varepsilon], [w]) \st w\in L \}$, and the $\widehat{\Sigma}$-driven transition function $\Delta \colon Q \times \Sigma \times (\Gamma^+ \cup \{\bot\}) \rightarrow Q \times (\Sigma \cup \{\varepsilon\}) \times (\Gamma^+ \cup \{\bot\})$, where $\Gamma=\Sigma \times Q$, is defined as follows:
	$\Delta$ maps $(([u], [v]), a, \stack{b,([u'],[v'])}\theta)$ to $(([a],[\varepsilon]), \varepsilon,\stack{a, ([u], [v])}\stack{b,([u'],[v'])}\theta)$ if $b \push a$,
	it returns $(([uva],[\varepsilon]), \varepsilon,\stack{a,([u'],[v'])}\theta)$ if $b\shift a$,
	and $(([u'],[v'uv]), a,\theta)$ if $b \pop a$.
	The soundness of our construction is given by the proof of the following lemma in Appendix.

%	\rednote{tom: add a 2-sentence summary about the proof, otherwise section 3.2 looks strange (or trivial).}
	\begin{lemma} \label{lem:recognizability}
		For every $L \subseteq \widehat{\Sigma}^*$, if $\equiv_L$ has finite index then $L$ is a $\widehat{\Sigma}$-OPL.
	\end{lemma}

	\section{Antichain-based Inclusion Checking}
	
	Considering two languages $L_1$ and $L_2$ given by some automata, the classical approach for deciding whether $L_1\subseteq L_2$ holds is to first compute the complement $\overline{L}_2$ of $L_2$, and then decide the emptiness of $L_1 \cap \overline{L}_2$.
	The major drawback with this approach is that the complementation requires the determinization of the automaton denoting $L_2$.
	A way to avoid the determinization is to search among words of $L_1$ for a counterexample to $L_1 \subseteq L_2$.
	For this, a breadth-first search can be performed symbolically as a fixpoint iteration.
	In order to guarantee its termination, the search is equipped with a well quasi-order, and considers only words that are not subsumed, i.e., the minima of $L_1$ with respect to the quasi-order.
	It is known that well quasi-orders satisfy the finite basis property, i.e., all sets of words have finitely many minima.
	Our approach is inspired by~\cite{DBLP:journals/tocl/GantyRV21} which, in the context of unstructured words, presents the antichain approach as a Galois connection, and observes that the upward closure of the quasi-order is a complete abstraction of concatenation according to the standard notion of completeness in abstract interpretation~\cite{DBLP:conf/popl/CousotC77}.
	We identify, in the context of structured words, sufficient conditions on quasi-orders to enable the antichain approach, by defining the class of \emph{language abstraction} quasi-orders (which satisfy the finite basis property).
	Further, we relax the syntactic congruence into a quasi-order that is a language abstraction of a given OPL.
	In particular, we prove that the syntactic congruence itself is a language abstraction for its language.
 	Then, we design our inclusion algorithm based on a fixpoint characterization of OPLs, which allows us to iterate breadth-first over all words accepted by a given OPA.
	Once equipped with a language abstraction quasi-order, this fixpoint is guaranteed to terminate, thus to synthesize a finite set $T\subseteq L_1$ of membership queries for $L_2$ which suffices to decide whether $L_1 \subseteq L_2$ holds.

	\subsection{Language Abstraction by Quasi-order}
	
	Let $E$ be a set of elements and $\preccurlyeq$ be a binary relation over $E$.
	The relation $\preccurlyeq$ is a \emph{quasi-order} when it is reflexive and transitive.
	A quasi-order $\preccurlyeq$ over $E$ is \emph{decidable} if for all $x, y \in E$, determining whether $x \preccurlyeq y$ holds is computable.
	Given a subset $X$ of $E$, we define its \emph{upward closure} with respect to the quasi-order $\preccurlyeq$ by ${_\preccurlyeq}{\upharpoonleft} X = \{e \in E \st \exists x \in X, x \preccurlyeq e\}$.
	Given two subsets $X, Y \subseteq E$ the set $X$ is a \emph{basis} for $Y$ with respect to $\preccurlyeq$, denoted $\mathfrak{B}(X \preccurlyeq Y)$, whenever
	$X \subseteq Y$ and ${_\preccurlyeq}{\upharpoonleft} X = {_\preccurlyeq}{\upharpoonleft} Y$.
	The quasi-order $\preccurlyeq$ is a \emph{well quasi-order} if and only if for each set $Y \subseteq E$ there exists a finite set $X \subseteq E$ such that $\mathfrak{B}(X \preccurlyeq Y)$.
	This property on bases is also known as the \emph{finite basis property}.
	Other equivalent definitions of well quasi-orders can be found in the literature~\cite{DBLP:journals/acta/LucaV94}, 
	we will use the following two:
	\begin{description}
		\item[$(\dagger)$] For every sequence $\{e_i\}_{i\in \N}$ in $E$, there exists $i, j \in \N$ with $i < j$ such that $e_i \preccurlyeq e_j$.
		\item[$(\ddagger)$] There is no sequence $\{X_i\}_{i\in\N}$ in $2^E$ such that ${_\preccurlyeq}{\upharpoonleft} X_1 \subsetneq {_\preccurlyeq}{\upharpoonleft} X_2 \subsetneq \dots$ holds.
	\end{description}

		%Throughout the section, we consider the OPAs $\A$ and $\B$ over $\widehat{\Sigma}_{cr}$ of Figures~\ref{A} and~\ref{B} to illustrate our definitions.
		%Clearly, we have $L(\B) \nsubseteq L(\A)$ as witnessed by $rc\in\widehat{\Sigma}^*$.

		Let $L_1, L_2$ be two languages.
	The main idea behind our inclusion algorithm is to compute a finite subset $T$ of $L_1$, called a \emph{query-basis}, such that $T\subseteq L_2 \Leftrightarrow L_1 \subseteq L_2$.
	Then, $L_1 \subseteq L_2$ holds if and only if each word of $T$ belongs to $L_2$, which is checked via finitely many membership queries.
%		In the sequel, we show how such a query-basis $T$ can be computed by the use of a quasi-order to prune the search of non-inclusion candidates in $S_1$.
	The computation of a query-basis consists of collecting enough words of $L_1$ to obtain a finite basis $T$ for $L_1$ with respect to a quasi-order $\preccurlyeq$ that abstracts $L_2$.
	%The computation of a query-basis for deciding whether $S_1$ is a subset of $S_2$ consists of collecting enough words of $S_1$ in order to obtain a finite basis $T$ for $S_1$ with respect to a ``well chosen'' quasi-order $\preccurlyeq$.
	When $\preccurlyeq$ is a well quasi-order, some basis is guaranteed to exist thanks to the finite basis property.
	%Given a quasi-order $\preccurlyeq$ over $\widehat{\Sigma}^*$, a word $w$ \emph{subsumes} another word $w'$ with respect to $\preccurlyeq$ if and only if $w \preccurlyeq w'$.
	To ensure the equivalence $L_1 \subseteq L_2 \Leftrightarrow T \subseteq L_2$ for any $T$ such that $\mathfrak{B}(T \preccurlyeq L_1)$, a counterexample $w \in L_1 \setminus L_2$ can be discarded (not included in $T$), only if it there exists $w_0\in T$ such that $w_0 \preccurlyeq w$ and $w_0$ is also a counterexample.
	Thus, we introduce the \emph{language saturation} property asking a quasi-order $\preccurlyeq$ to satisfy the following: for all $w_0, w \in\widehat{\Sigma}^*$ if $w_0 \preccurlyeq w$ and $w_0 \in L_2$ then $w\in L_2$, or equivalently, ${_\preccurlyeq}{\upharpoonleft}L_2 = L_2$.
	Intuitively, language saturation ensures the completeness of the language abstraction with respect to the inclusion.
	Finally, to guarantee that the query-basis $T$ is iteratively constructible with an effective fixpoint computation, the quasi-order $\preccurlyeq$ must be both chain-monotonic and decidable.
	We now define the notion of \emph{language abstraction} to identify the properties for a quasi-order over structured words that allow an effectively computable query-basis, as was done in~\cite{DBLP:conf/cav/DoveriGM22,DBLP:journals/tocl/GantyRV21} in the context of B\"uchi automata for quasi-orders over unstructured infinite words.

	\begin{definition}[language abstraction]
		Let $L \subseteq \widehat{\Sigma}^*$.
		A quasi-order $\preccurlyeq$ over $\widehat{\Sigma}^*$ is a \emph{language abstraction} of $L$ iff (1) it is decidable, (2) it is chain-monotonic, (3) it is a well quasi-order, and (4) it saturates $L$. %, i.e., if $w_0 \preccurlyeq w$ and $w_0 \in L$ then $w\in L$, for all $w_0, w \in\widehat{\Sigma}^*$.
	\end{definition}
	
	In the next section, we provide an effective computation of a query-basis for an OPA, thanks to a quasi-order that abstracts its language.

	%I THINK WHAT IS ABOVE THE DEFINITION SUFFICE
	%Notice that we require language abstractions to be well quasi-orders. i.e., to have the finite basis property. \rednote{more on finiteness? explain this?}

	\begin{figure}[t!]
		\begin{minipage}{\linewidth}
			\centering
			
			\begin{minipage}{.35\linewidth}\centering
				\includegraphics[scale=0.5]{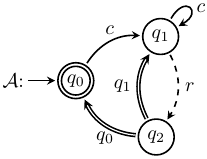}
%				\scalebox{0.75}{
%					\begin{tikzpicture}[node distance=2cm, initial text = $\A$:]
%						\node (nodecenter) {};
%						\node[state, initial, label=center:$q_0$, shift=(175:1cm), final] (q0) at (nodecenter) {};
%						\node[state, label=center:$q_1$, shift=(60:1cm)] (q1) at (nodecenter) {};
%						\node[state, label=center:$q_2$, shift=(295:1cm)] (q2) at (nodecenter) {};
%						\path[transition]
%						(q0) edge[bend left] node[above] {$c$} (q1)
%						(q1) edge[loop above right] node[right] {$c$} (q1)
%						(q1) edge[bend left, dashed] node[right] {$r$} (q2)
%						(q2) edge[bend left, double] node[left] {$q_1$} (q1)
%						(q2) edge[bend left, double] node[below] {$q_0$} (q0)
%						;	
%					\end{tikzpicture}
%				}
			\end{minipage}
			\begin{minipage}{.25\linewidth}\centering
				\scalebox{.9}{
				\bgroup
				\renewcommand{\arraystretch}{.95}
				\setlength{\arraycolsep}{2pt}
				$\begin{array}{c|ccc}
					\widehat{\Sigma}_{cr} &   c   &    r   & \varepsilon \\\hline
					c      & \push & \shift & \pop        \\
					r      & \pop  &  \pop  & \pop        \\
					\varepsilon & \push & \push  & \shift      \\
				\end{array}$
				\egroup}
			\end{minipage}
			\begin{minipage}{.38\linewidth}
				\includegraphics[scale=0.5]{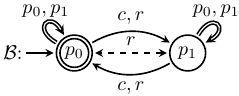}
%				\scalebox{0.75}{
%					\begin{tikzpicture}[node distance=2cm, initial text = $\B$:]
%						\node[state, initial, label=center:$p_0$, final] (p0) {};
%						\node[state, label=center:$p_1$, right of = p0] (p1) {};
%						\path[transition]
%						(p0) edge[bend left] node[above] {$c, r$} (p1)
%						(p1) edge[bend left] node[below] {$c, r$} (p0)
%						(p1) edge[<->, dashed] node[above] {$r$} (p0)
%						(p0) edge[loop above left, double] node[above] {$p_0, p_1$} (p0)
%						(p1) edge[loop above right, double] node[above] {$p_0, p_1$} (p1)
%						;
%					\end{tikzpicture}
%				}
			\end{minipage}

			\begin{minipage}[t]{.45\linewidth}
				\caption{(left) OPA $\A$ over $\widehat{\Sigma}_{cr}$ recognizing the VPL of well-matched \emph{call}/\emph{return} words.}\label{A}
			\end{minipage}
			\hfill
			\begin{minipage}[t]{.45\linewidth}
				\caption{(right) OPA $\B$ over $\widehat{\Sigma}_{cr}$ recognizing the regular language of words of even length.}\label{B}
				\label{table:inclusion}
			\end{minipage}
		\end{minipage}
	\end{figure}
	
	\begin{example}\label{example:inclusion}
		The operator precedence alphabet $\widehat{\Sigma}_{cr}$ of $\A$ and $\B$ from Figures~\ref{A} and~\ref{B} induces four families of words:
		(1) the words of $\widehat{\Sigma}^*_{\shift}$ where every $c$ matches an $r$,
		(2) the words of $\widehat{\Sigma}^*_{\push} = \widehat{\Sigma}^*_{\pusheq} \setminus \widehat{\Sigma}^*_{\shift}$  where some $c$ is pending for an $r$ on its right,
		(3) the words of $\widehat{\Sigma}^*_{\pop} = \widehat{\Sigma}^*_{\popeq} \setminus \widehat{\Sigma}^*_{\shift}$ where some $r$ is pending for a $c$ on its left, 
		and (4) all other words of $\widehat{\Sigma}^*_{\nshift} = \Sigma^* \setminus \left(\widehat{\Sigma}^*_{\pusheq} \cup \widehat{\Sigma}^*_{\popeq} \right)$.
		
		We focus on deciding whether $L(\B)$ is a subset of $L(\A)$ and suppose that we are given the quasi-order $\ll$ that is a language abstraction of $L(\A)$.
		Additionally, we have that two words compare with $\ll$ only if they belong to the same family, and we have the following bases: $\mathfrak{B}(\{cr\} \ll \widehat{\Sigma}^*_{\shift})$, $\mathfrak{B}(\{c\} \ll \widehat{\Sigma}^*_{\push})$, $\mathfrak{B}(\{r\} \ll \widehat{\Sigma}^*_{\pop})$, and $\mathfrak{B}(\{rc\} \ll \widehat{\Sigma}^*_{\nshift})$.
		We observe that $\ll$ saturates $L(\A)$ since $\widehat{\Sigma}^*_{\shift} \subseteq L(A)$ and $\widehat{\Sigma}^*_{\pusheq}, \widehat{\Sigma}^*_{\popeq}, \widehat{\Sigma}^*_{\nshift} \nsubseteq L(\A)$.
		
		Among the representatives $cr$, $c$, $r$, and $rc$, we can construct the set $T = \{cr, rc\}$ since $c, r \notin L(\B)$.
		The set $T$ is a query-basis for deciding whether $L(\B)$ is a subset of $L(\A)$.
		In particular, $rc \in T$ witnesses that $L(\B) \nsubseteq L(\A)$.
	\end{example}
	
	Note that the syntactic congruence is a natural language abstraction of OPLs.
	
	\begin{proposition}
		For every OPL $L$, ${\equiv_{L}}$ is a language abstraction of $L$.
	\end{proposition}

	When the language to be abstracted is given by an OPA we are able to define a quasi-order, called \emph{structural quasi-order}, that is 
	based on the underlying structure of the automaton.

	\begin{definition}[structural quasi-order]
		Given an OPA $\A=(Q, I, F, \Delta)$, we define the relation $\leqslant_\A$ over $\widehat{\Sigma}^*$ as follows:
		\[x \leqslant_{\A} y \iff x \approx y \land \forall q, q' \in Q, \forall \gamma \in \Gamma \cup \{\bot\} \bigwedge \begin{cases}
			f_{x}(q,\gamma) \subseteq f_{y}(q,\gamma)\\
			g_{x}(q,q',\gamma) \subseteq g_{y}(q,q',\gamma)\\
			(\varPhi_{x}(q,\gamma))^\top \subseteq (\varPhi_{y}(q,\gamma))^\top
		\end{cases}\]
	\end{definition}

	\begin{remark}\label{remark:relax} \hspace{-0.16cm}
		For every OPA $\A$, the quasi-order ${\leqslant_{\A}}$ relaxes the congruence ${\equiv_A}$ from Section~3.\\
		For every OPA $\A$, the quasi-order ${\leqslant_{\A}}$ relaxes the congruence ${\equiv_A}$ from Section~3.
%		Clearly ${\leqslant_{\A}}$ relaxes ${\equiv_A}$ from Section~3, i.e., $\equiv_{\A} \implies \leqslant_{\A}$.
	\end{remark}
	
	Note that, for every OPA $\A$, the set $Q \times (\Gamma \cup \{\bot\})$ is finite.
	Consequently, $\leqslant_\A$ is computable, and it is a well quasi-order since there cannot exist an infinite sequence of incomparable elements, i.e., ($\dagger$) holds.
	
	\begin{proposition}\label{proposition:monotonicity}
		For every OPA $\A$, $\leqslant_\A$ is a computable chain-monotonic well quasi-order.
	\end{proposition}
		
	Next, we establish that structural quasi-orders saturate their languages.
	
	\begin{lemma}\label{lemma:preserve}
		For every OPA $\A$ and $w_1, w_2 \in \widehat{\Sigma}^*$, if $w_1 \leqslant_\A  w_2$ and $w_1 \in L(\A)$ then $w_2 \in L(\A)$.
	\end{lemma}

	The following comes as a direct consequence of Proposition~\ref{proposition:monotonicity} and Lemma~\ref{lemma:preserve}.
	
	\begin{corollary}
		For every OPA $\A$, $\leqslant_\A$ is a language abstraction of $L(\A)$.
	\end{corollary}

	We continue Example~\ref{example:inclusion}, showing that the structural quasi-order agrees with the considered bases above.

	%Building on Example~\ref{example:inclusion}, we demonstrate that a structural quasi-order is indeed a language abstraction.\greennote{to rephrase}
	
	\begin{example}
		The quasi-order $\ll$ described in Example~\ref{example:inclusion} agrees with the structural quasi-order $\leqslant_{\A}$ of the OPA $\A$ in Figure~5.
		Indeed, due to the constraint that two comparable words $x,y \in \widehat{\Sigma}^*$ should be chain equivalent, i.e., $x \approx y$, % equality constraints on heads and tails,
		the quasi-order $\leqslant_{\A}$ compares only the words from the same families among $\widehat{\Sigma}^*_{\shift}, \widehat{\Sigma}^*_{\push}, \widehat{\Sigma}^*_{\pop}$, and $\widehat{\Sigma}^*_{\nshift}$.
		We also note that, for all words, adding a factor in $\widehat{\Sigma}^*_{\shift}$ cannot change the accessibility in $\A$ since reading such a factor has no effect on the stack or the current state.
		Additionally, reading several $c$ in a row triggers a self loop and reading several $r$ is not possible in $\A$.
		As a consequence, the base predicates mentioned in Example~\ref{example:inclusion} hold, that is, $\mathfrak{B}(\{cr\} \leqslant_{\A} \widehat{\Sigma}^*_{\shift})$, $\mathfrak{B}(\{c\} \leqslant_{\A} \widehat{\Sigma}^*_{\push})$, $\mathfrak{B}(\{r\} \leqslant_{\A} \widehat{\Sigma}^*_{\pop})$, and $\mathfrak{B}(\{rc\} \leqslant_{\A} \widehat{\Sigma}^*_{\nshift})$.
		Yet, we have that $cr \leqslant_{\A} \varepsilon$ because $(q_0, cr, \bot) \tikzpath (q_2, \varepsilon, \stack{c, q_0})$ but $(q_0, \varepsilon, \bot)\tikznpath  (q_2,\varepsilon, \stack{c, q_0})$.
	\end{example}

	\subsection{Fixpoint Characterization of Languages and Inclusion}
	
		In order to formulate our inclusion algorithm, it remains to give an effective computation of a query-basis.
		We do so through a fixpoint characterization of the languages recognized by OPAs.
		We introduce the function $\texttt{Cat}$ to construct words that follow the runs of the given OPA.
		Iterating the $\texttt{Cat}$ function $n\in\N$ times captures all words of length up to $n$, and the fixpoint of the iteration captures the entire language of a given OPA.

		%We start with the fixpoint characterization of the languages recognized by a given OPA.
		Let $\A = (Q, I, F, \Delta)$ be an OPA.
		Consider a vector of set of words $\vec{X}$ that accesses its fields with two states $s, t \in Q$, and three letters $a, b, c \in \widehat{\Sigma} \cup \{\varepsilon\}$.
		Intuitively, we aim at constructing $\vec{X}$ iteratively such that, reading any $w \in \vec{X}^{a, b, c}_{s, t}$ from the configuration $(s, wc, \alpha)$ where $\alpha^{\top}=a$ allows reaching  $(t, c, \beta)$ where $\beta^{\top}=b$ in $\A$.
		We recall that $\bot^{\top}=\varepsilon$.
		As the base case, we take $\vec{X}^{a, b,c}_{s, t} = \varepsilon$ when $a=b$ and $s=t$, otherwise $\vec{X}^{a, b,c}_{s, t} = \varnothing$.
		Then, we introduce operations (more explicitly, functions from sets of words to sets of words) that use the transitivity of ${\tikzpath}$ in $\A$ to extend the sets of $\vec{X}$.
		We first introduce:
		\[
			\texttt{CatShift}(\vec{X}^{a, b, c}_{s, t}) =
			\left\{
			\begin{array}{c@{~}|@{~}c}
				u b' v &
				\begin{array}{l}
					a',b' \in \Sigma, \,
					q, s',t'\in Q, \,
					u \in \vec{X}^{a, a', b'}_{s, s'}, \,
					v\in \vec{X}^{b', b, c}_{t', t}, \\
					(s', \stack{a', q}\bot) \tikzshift{b'} (t', \stack{b', q}\bot)
				\end{array}
			\end{array}
			\right\}
		\]
		Essentially, $\texttt{CatShift}$ adds $ub'v$ to $\vec{X}^{a, b, c}_{s, t}$ when some run over $u$ can be appended with $b'$ thanks to a shift-transition, 
		and some run of $v$ requires starting with $b'$ at the top of the stack.
		Next, we introduce:
		\[
			\texttt{CatChain}(\vec{X}^{a, b, c}_{s, t}) = 
			\left\{
			\begin{array}{c@{~}|@{~}c}
				u b' v &
				\begin{array}{l}
					a',b',c' \in \Sigma, \,
					q, s',t'\in Q, \,
					u \in \vec{X}^{a, b, b'}_{s, q}, \,
					v\in \vec{X}^{b', c', c}_{s', t'}, \\
					b \push b' \, \land \, (q, \bot) \tikzpush{b'} (s', \stack{b', q}\bot) \, \land \,
					(t', \stack{c', q}\bot) \tikzpop{c} (t, \bot)
				\end{array}
			\end{array}
			\right\}
		\]
		Intuitively, $\texttt{CatChain}$ adds $ub'v$ to $\vec{X}^{a, b, c}_{s, t}$ when some run over $u$ can be appended with $b'$ thanks to a push-transition, and some run of $v$ requires starting with $b'$ at the top of the stack.
		Additionally, $b'$ is guaranteed to be removed from the stack thanks to a pop-transition on the incoming letter $c$.
		Finally, we define: 
		\[
			\texttt{Cat}(\vec{X}^{a, b, c}_{s, t}) = 
			\vec{X}^{a, b, c}_{s, t} \cup
			\texttt{CatShift}(\vec{X}^{a, b, c}_{s, t}) \cup 
			\texttt{CatChain}(\vec{X}^{a, b, c}_{s, t})
		\]
		Note that the function $\texttt{Cat}$ never removes words from the sets of  $\vec{X}$, i.e., $\vec{X}^{a, b, c}_{s, t} \subseteq \texttt{Cat}(\vec{X}^{a, b, c}_{s, t})$.
		Iterating the $\texttt{Cat}$ function $n\in\N$ times allows us to extend the sets of $\vec{X}$ to words of length at most $n$ that follow some run of $\A$.
		In particular, $\texttt{Cat}$ characterizes the language of $\A$ by $w\in L(\A)$ if and only if $w\in \texttt{Cat}^*(\vec{X}^{\varepsilon, \varepsilon, \varepsilon}_{q_I, q_F})$ for some $q_I\in I$ and $q_F\in F$.
		This is formalized by the following lemma.

		\begin{lemma}\label{lemma:cat}
			Let $\A=(Q, I, F, \Delta)$ be an OPA, and let $\Gamma=\Sigma \times Q$.
			Considering $\vec{U}^{a, b,c}_{s, t} = \varepsilon$ when $a=b$ and $s=t$, otherwise $\vec{U}^{a, b,c}_{s, t} = \varnothing$.
			The following holds for all $n>0$:
			\[\texttt{Cat}^n(\vec{U}^{a, b, c}_{s, t}) {=} \big\{ u \mid
				(s, uc, \alpha){\tikzpath}(t, c, \beta),
				|u|=n, 
				\alpha\in\varTheta_a, \beta\in\varTheta_b,
				au\in\widehat{\Sigma}^*_{\pusheq},
				uc\in\widehat{\Sigma}^*_{\popeq}, u^{\last}=b
			\big\}\]
			where, for all $a\in\widehat{\Sigma}$, the set of stack symbols $\varTheta_a \subseteq \Gamma \cup \{\bot\}$ is defined by $\varTheta_a=\{\bot\}$ if $a=\varepsilon$, and $\varTheta_a=\{\stack{a, q} \st q\in Q\}$ otherwise.
		\end{lemma}

		We continue Example~\ref{example:inclusion}, showing that $\texttt{Cat}$ agrees with the considered query-basis.

		\begin{example}
			Let $\vec{U}^{a, b,c}_{s, t} = \varepsilon$ when $a=b$ and $s=t$, otherwise $\vec{U}^{a, b,c}_{s, t} = \varnothing$.
			Thanks to Lemma~\ref{lemma:cat}, we have that $L(\B)=\texttt{Cat}^*(\vec{U}^{\varepsilon, \varepsilon, \varepsilon}_{p_0, p_0})$.
			First observe that $c, r \notin \texttt{Cat}^*(\vec{U}^{\varepsilon, \varepsilon, \varepsilon}_{p_0, p_0})$.
			This comes from Lemma~\ref{lemma:cat} and the fact that there is no run of $\B$ from $p_0$ to $p_0$ that reads a single letter.
			Next, we prove that $cr, rc \in \texttt{Cat}^{2}(\vec{U}^{\varepsilon, \varepsilon, \varepsilon}_{p_0, p_0})$.
			
			We show that $r\in \texttt{Cat}(\vec{U}^{\varepsilon, \varepsilon, c}_{p_0, p_1})$ by $\texttt{CatChain}$.
			Indeed, we have
				$\varepsilon\in\vec{U}^{\varepsilon, \varepsilon, r}_{p_0, p_0}$,
				$\varepsilon\in\vec{U}^{r, r, c}_{p_1, p_1}$,
				$\varepsilon \push r$,
				and $(p_0, \bot) \tikzpush{r} (p_1, \stack{r, p_1}\bot) \tikzpop{c} (p_1, \bot)$.
			Then, $rc \in \texttt{Cat}^{2}(\vec{U}^{\varepsilon, \varepsilon, \varepsilon}_{p_0, p_0})$ by $\texttt{CatChain}$ since
				$r\in \texttt{Cat}(\vec{U}^{\varepsilon, \varepsilon, c}_{p_0, p_1})$, 
				$\varepsilon\in\vec{U}^{c, c, \varepsilon}_{p_0, p_0}$, 
				$\varepsilon \push c$, 
				and $(p_1, \bot) \tikzpush{c} (p_0, \stack{c, p_1}\bot) \tikzpop{\varepsilon} (p_1, \bot)$.

			We show that $r\in \texttt{Cat}(\vec{U}^{c, r, \varepsilon}_{p_1, p_0})$ by $\texttt{CatShift}$.
			Indeed, we have
				$\varepsilon\in\vec{U}^{c, c, r}_{p_1, p_1}$,
				$\varepsilon\in\vec{U}^{r, r, \varepsilon}_{p_0, p_0}$,
				and $(p_1, \stack{c, p}\bot) \tikzshift{r} (p_0, \stack{r, p}\bot)$, for all $p\in\{p_0, p_1\}$.
			Then, $cr \in \texttt{Cat}^{2}(\vec{U}^{\varepsilon, \varepsilon, \varepsilon}_{p_0, p_0})$ by $\texttt{CatChain}$ since
				$\varepsilon\in\vec{U}^{\varepsilon, \varepsilon, c}_{p_0, p_0}$,
				$r\in \texttt{Cat}(\vec{U}^{c, r, \varepsilon}_{p_1, p_0})$, 
				$\varepsilon \push c$,
				$(p_0, \bot) \tikzpush{c} (p_1, \stack{c, p_0}\bot)$,
				and $(p_0, \stack{r, p_0}) \tikzpop{\varepsilon} (p_0, \bot)$.
		\end{example}

		The computation of a query-basis for deciding whether $L_1$ is a subset of $L_2$ consists of iterating $\texttt{Cat}$ to collect enough words to obtain a vector of finite bases with respect to the quasi-order $\preccurlyeq$ that is a language abstraction of $L_2$.
		In other words, we search for $n\in \N$ such that $\texttt{Cat}^n(\vec{X}^{a, b, c}_{s, t})$ is a basis for $\lim_{k \mapsto \infty} \texttt{Cat}^{k}(\vec{U}^{a, b, c}_{s, t})$ with respect to $\preccurlyeq$.
		The following lemma shows that when $\mathfrak{B}(\texttt{Cat}^n(\vec{X}^{a, b, c}_{s, t}) \preccurlyeq \texttt{Cat}^{n+1}(\vec{X}^{s, b, c}_{s, t}))$ holds for some $n\in\N$, then $\mathfrak{B}(\texttt{Cat}^n(\vec{X}^{a, b, c}_{s, t}) \preccurlyeq \lim_{k \mapsto \infty}\texttt{Cat}^{k}(\vec{X}^{a, b, c}_{s, t}))$ holds also, as long as the used quasi-order is chain-monotonic.

		\begin{lemma}\label{lemma:base}
			Let $\preccurlyeq$ be a chain-monotonic quasi-order over $\widehat{\Sigma}^*$.
			For every $A = (Q, I, F, \Delta)$ and $\vec{X}, \vec{Y}$ such that $\mathfrak{B}(\vec{X}^{a, b, c}_{s, t} \preccurlyeq \vec{Y}^{a, b, c}_{s, t})$ holds for all $s, t \in Q$ and all $a, b, c\in\Sigma \cup \{\varepsilon\}$, we have $\mathfrak{B}(\texttt{Cat}(\vec{X}^{a, b, c}_{s, t}) \preccurlyeq \texttt{Cat}(\vec{Y}^{a, b, c}_{s, t}))$ holds also for all $s, t \in Q$ and all $a, b, c\in\Sigma \cup \{\varepsilon\}$.
		\end{lemma}

	\begin{figure}[ht!]\centering
	\centering
%	\begin{tikzpicture}[scale=0.9]
%			\node[text width=\linewidth-3em]{\hspace{-0em}
				\begin{minipage}{\linewidth}
					\begin{algorithm}[H]
						\DontPrintSemicolon
						\SetInd{.5em}{1em}
						\SetKwIF{If}{ElseIf}{Else}{if}{then}{else if}{else}{}
						\SetKwFor{For}{for each}{do}{}
						\SetKwIF{DefIf}{DefElseIf}{DefElse}{}{if}{}{else}{}
						\SetKwFor{Def}{let}{as}{}
						\SetKwRepeat{Repeat}{repeat}{until}
						\SetKw{Return}{return}
						\SetKw{And}{and}
						\SetKwProg{FUNCTON}{Function}{:}{}
						\SetKwInput{IN}{Input}
						\SetKwInput{OUT}{Output}
						\SetKwInput{DATA}{Data}
						\SetKwComment{ccc}{\color{gray}/* }{\color{gray}\ */\quad}
						\IN{an OPL $L_1$ given by the OPA $(Q, I, F, \Delta)$}
						\IN{a language $L_2$ with a procedure deciding if $w \in L_2$}
						\IN{a quasi-order $\preccurlyeq$ that is a language abstraction of $L_2$} %$(w_0 \preccurlyeq w \land w_0 \in S) \implies w \in S$}
					\OUT{Returns \texttt{ok} if $L_1\subseteq L_2$ and \texttt{ko} otherwise}
					\BlankLine
					
					\SetAlgoNoLine
					\FUNCTON{}{
						%						\SetAlgoVlined
						\lDef*{$\vec{U}$}{\leDefIf*{$\vec{U}^{a, b, c}_{s, t} \coloneqq \varepsilon$}{$a=b \land s=t$}{$\vec{U}^{a, b, c}_{s, t} \coloneqq \varnothing$}}
						
						$\vec{X} \coloneqq \vec{U}\)%\nllabel{algo:forq:line:Winit}\;
						
						\Repeat
						{$\mathfrak{B}(\vec{X}^{a, b, c}_{s, t} \preccurlyeq \texttt{Cat}(\vec{X}^{a, b, c}_{s, t}))$ for all $s, t \in Q$ and all $a, b, c \in \Sigma\cup\{\varepsilon\}$}
						{\lDef*{$\vec{X}$}{$\vec{X}^{a, b, c}_{s, t} \coloneqq \texttt{Cat}(\vec{X}^{a, b, c}_{s, t})$}}%\nllabel{algo:forq:line:loopU}
						%$\vec{X}^{a, b, c}_{s, t} \coloneqq \texttt{Cat}(\vec{X}^{a, b, c}_{s, t})$ \qquad for all $s, t \in Q$ and all $a, b, c \in \Sigma\cup\{\varepsilon\}$}} \nllabel{algo:forq:line:loopU}

				\For{$(q_I, q_F) \in I \times F$}{%\nllabel{algo:forq:line:final}
					\For{$w\in \vec{X}^{\varepsilon, \varepsilon, \varepsilon}_{q_I, q_F}$}{%\nllabel{algo:forq:line:U}
						\lIf{$w\notin L_2$}{\Return ko}}%\nllabel{algo:forq:line:ko}
				}
				\Return ok}
		\end{algorithm}
	\end{minipage}
%	\end{tikzpicture}
	\caption{Antichain inclusion algorithm.}\label{algo}
	\end{figure}
	Our inclusion algorithm is given in \figurename~\ref{algo}.
	We can prove that it always terminates thanks to the finite base property of language abstractions.
	Additionally, its correctness is based on the following: Lemmas~\ref{lemma:cat} and~\ref{lemma:base} ensure that the repeat-until loop computes a basis of the language $L_1$ given by an OPA while the language saturation ensures the completeness of this basis with respect to the inclusion problem.
	
	\begin{theorem}\label{theorem:correct}
		The algorithm from \figurename~\ref{algo} terminates and decides language inclusion.
	\end{theorem}

	We establish that our inclusion algorithm for OPAs is in \textsc{ExpTime} as a consequence of Lemma~\ref{lem:finite}, Remark~\ref{remark:relax}, the facts that the vector $\vec{X}$ maintains polynomially many sets of words and the membership problem for OPAs is in \textsc{PTime} (Remark~\ref{rem:OPAcomplexity}).
	We recall that inclusion and universality are \textsc{ExpTime-C} for both OPLs and VPLs \cite{DBLP:conf/stoc/AlurM04,DBLP:journals/siamcomp/LonatiMPP15}.
	%\rednote{tom: please add an explicit exponential upper bound on the complexity to thm 40, or in the text after, as a O() expression that shows the parameters (number of states etc.) in the exponent (probably only parameters of B, not A)}

	\begin{theorem}\label{theorem:complexity}
		For all OPAs $\A, \B$ with respectively $n_{\A}, n_{\B}$ states and $m$ input letters, the inclusion algorithm from \figurename~\ref{algo} with ${\leqslant_\B}$ as the language abstraction quasi-order decides if $L(\A) \subseteq L(\B)$ in time
		$\mathcal{O}(m\times n_{\A})^{\mathcal{O}(m \times n_{\B})^{\mathcal{O}(1)}}$.
		%$\mathcal{O}(|\Sigma|^3\times n_{\A}^2 \times n_{\B})^{\mathcal{O}(|\Sigma|\times|n_{\B}|)^{\mathcal{O}(1)}}$.
	\end{theorem}

	\section{Conclusion}
	
	We provided, for the first time, a syntactic congruence that characterizes operator precedence languages (OPLs) in the following exact sense:
	for any language $L$, the syntactic congruence has finitely many equivalence classes if and only if $L$ is an OPL.
	Second, we gave sufficient conditions for a quasi-order to yield an antichain algorithm for solving the universality and language inclusion problems for nondeterministic automata. 
	These conditions are satisfied by our syntactic congruence, which, like any finite congruence, is monotonic for structured words (i.e., chain-monotonic) and saturates its language.
	%We then showed that our syntactic congruence (indeed, any finite syntactic congruence) satisfies these conditions,
	This results in an exponential-time antichain algorithm for the inclusion of operator precedence automata (OPAs),
	which is the optimal worst-case complexity for the \textsc{ExpTime}-hard problem.
	This will allow efficient symbolic implementations of antichain algorithms to be extended to OPLs.
	
	The possibility of future research directions regarding OPLs is still vast.
	One promising direction is to study OPAs from a runtime verification~\cite{DBLP:series/lncs/10457} perspective.
	For example, extending the runtime approaches for visibly pushdown automata~\cite{DBLP:conf/rv/BruyereDG13,DBLP:conf/rv/RosuCB08},
	one can study the monitor synthesis and right-universality problems for OPAs to establish them as an expressively powerful class of monitors.
	Also other methods developed for visibly pushdown automata may be generalizable to OPAs based on our syntactic congruence,
	such as learning algorithms~\cite{DBLP:conf/concur/KumarMV06}.
	
	While OPLs characterize the weakest known restrictions on stack operations which enable decidability of the inclusion problem, 
	one may try to push the frontier of decidability by relaxing the restrictions on stack operations further.
	Investigating similar restrictions in the context of observability for counter automata can also provide new decidability results.
	For example, \cite{DBLP:conf/fsttcs/Bollig16} shows that hardcoding the counter operations (increments and decrements) in the input letters yields decidable inclusion for one-counter automata.
	Another natural direction is to investigate quantitative versions of OPAs,
	for instance, through the addition of Presburger acceptance constraints,
	and to identify decidable fragments thereof~\cite{DBLP:journals/iandc/DrosteDMP22}.

	\bibliography{icalp23_with_appendix}

	\newpage
	\appendix
	
	\section{Omitted Proofs}
	
	\paragraph*{Proposition~\ref{proposition:chain:finite}}
	\proofsubparagraph*{Statement.}
		$\approx$ is a chain-monotonic equivalence relation with finitely many classes.

	\begin{proof}
		The reflexivity, symmetry, and transitivity of ${\approx}$ is trivial, and the chain-monotonicity follows from the definitions.
		We prove here that ${\approx}$ has finitely many equivalence classes.
		
		Consider $w \in \widehat{\Sigma}^+$ for which $\lambda(w)$ is of the form $a_1\dots a_\ell b_1 \dots b_m c_1 \dots c_n$ for some $\ell, m, n\in \N$ and such that
		\[a_1 \popeq a_2 \popeq \ldots \popeq a_\ell \pop b_1 \shift b_2 \shift \ldots \shift b_m \push c_1\pusheq c_2 \pusheq \ldots \pusheq c_n\]
		where $a_i, b_j, c_k \in \Sigma$ for all $i, j, k$.
		Let $P_w^{\first} = \{a_1, b_1\} \cup \{a_{i+1} \st a_i \pop a_{i+1}\}$ and $P_w^{\last} = \{b_m, c_n\} \cup \{c_i \st c_i \push c_{i+1}\}$.
		By convention, we define $P_{\varepsilon}^{\first} = P_{\varepsilon}^{\last} = \{\varepsilon\}$.
		We can define the profile of a word as $P_w = (w^{\first}, w^{\last}, P_w^{\first}, P_w^{\last})$.
		Note that there are at most $|\widehat{\Sigma}|^{2} \times 2^{2|\widehat{\Sigma}|}$ distinct profiles.
		We show that $P_x = P_y$ implies $x \approx y$ for all $x, y \in \widehat{\Sigma}^*$.
		
		Let $x, y \in \widehat{\Sigma}^*$ be such that $P_x = P_y$.
		By the equality of profiles, we directly get $x^\first = y^\first$ and $x^\last = y^\last$.
		Now, let $u_0,v_0 \in \widehat{\Sigma}^*$ be such that $u_0x^{\first}\in \widehat{\Sigma}^*_{\popeq}$ and $x^{\last}v_0\in \widehat{\Sigma}^*_{\pusheq}$.
		Since $x$ and $y$ agree on the first and last letters, we $u_0$ and $v_0$ also satisfy $u_0y^{\first}\in \widehat{\Sigma}^*_{\popeq}$ and $y^{\last}v_0\in \widehat{\Sigma}^*_{\pusheq}$.
		Now, let $u,v \in \widehat{\Sigma}^*$ be arbitrary.
		We want to show that $\chain{u}{u_0 x v_0}{v}$ iff $\chain{u}{u_0 y v_0}{v}$.
		Note that $\lambda(x)$ is of the form given above, and since $u_0x^{\first}\in \widehat{\Sigma}^*_{\popeq}$ and $x^{\last}v_0\in \widehat{\Sigma}^*_{\pusheq}$, the word $u_0 x v_0$ is also of the same form.
		Moreover, since $y$ has the same profile as $x$, the same holds for $u_0 y v_0$ as well.
		In particular, $P_{u_0 x v_0} = P_{u_0 y v_0}$.
		Then, if $\chain{u}{u_0 x v_0}{v}$, we have $u^\last \push a$ for all $a \in P_{u_0 x v_0}^\first$ and $b \pop v^\first$ for all $b \in P_{u_0 x v_0}^\last$. 
		Since the same holds for $u_0 y v_0$, we get $\chain{u}{u_0 y v_0}{v}$.
		The other direction is similar.
	\end{proof}
	
	\paragraph*{Theorem~\ref{thm:congruencechainmono}}
	\proofsubparagraph*{Statement.}
		For every $L\subseteq \widehat{\Sigma}^*$, $\equiv_L$ is a chain-monotonic equivalence relation.

	\begin{proof}
		The reflexivity, symmetry, and transitivity of ${\equiv_L}$ is trivial.
		We prove that ${\equiv_L}$ is chain-monotonic.
		Let us define the following relation over $\widehat{\Sigma}^*$:
		\[
			x \sim y \iff
			\forall u, v, u_0, v_0 \in \widehat{\Sigma}^*
			\begin{cases}\begin{array}{c}	
					\big(u_0x^{\first}\in \widehat{\Sigma}^*_{\popeq} \land x^{\last}v_0\in \widehat{\Sigma}^*_{\pusheq} \land \chain{u}{u_0xv_0}{v}\big)\\
					\Downarrow\\
					\big(uu_0xv_0v\in L \Leftrightarrow uu_0yv_0v\in L\big)
				\end{array}
			\end{cases}
		\]
		Recall that for every $x, y \in \widehat{\Sigma}^*$ we have $x \equiv_L y$ if and only if both $x \approx y$ and $x \sim y$ holds, where ${\approx}$ is the chain equivalence.
	
		Let $x,y \in \widehat{\Sigma}^*$ such that $x \equiv_L y$, and let $u, v, u_0, v_0 \in \widehat{\Sigma}^*$ such that $u_0 x^{\first}, u_0 y^{\first} \in \widehat{\Sigma}^*_{\popeq}$, $x^{\last} v_0, y^{\last} v_0 \in \widehat{\Sigma}^*_{\pusheq}$, $\chain{u}{u_0 x v_0}{v}$, and $\chain{u}{u_0 y v_0}{v}$.
		We claim that $u u_0 x v_0 v \equiv_{L} u u_0 y v_0 v$.
		Since the chain equivalence $\approx$ is chain-monotonic, we get $u u_0 x v_0 v \approx u u_0 y v_0 v$.
		Then, we only need to show that $u u_0 x v_0 v \sim u u_0 y v_0 v$.	
		In particular, for every $u', v', u_0', v_0' \in \widehat{\Sigma}^*$ such that $u_0' (u u_0 x v_0 v)^{\first} \in \widehat{\Sigma}^*_{\popeq}$,  $(u u_0 x v_0 v)^{\last} v_0' \in \widehat{\Sigma}^*_{\pusheq}$, $\chain{u'}{u_0' u u_0 x v_0 v v_0'}{v'}$ we have $u' u_0' u u_0 x v_0 v v_0' v' \in L $ iff $u' u_0' u u_0 y v_0 v v_0' v' \in L$.
		
		Let $u', v', u_0', v_0' \in \widehat{\Sigma}^*$ be such that
		$u_0' (u u_0 x v_0 v)^{\first} \in \widehat{\Sigma}^*_{\popeq}$, 
		$(u u_0 x v_0 v)^{\last} v_0' \in \widehat{\Sigma}^*_{\pusheq}$,
		and $\chain{u'}{u_0' u u_0 x v_0 v v_0'}{v'}$.
		Furthermore, since $u u_0 x v_0 v \approx u u_0 y v_0 v$, we have
		$u_0' (u u_0 y v_0 v)^{\first} \in \widehat{\Sigma}^*_{\popeq}$, 
		$(u u_0 y v_0 v)^{\last} v_0' \in \widehat{\Sigma}^*_{\pusheq}$, and
		$\chain{u'}{u_0' u u_0 y v_0 v v_0'}{v'}$.
		
		Now, take $u_0'' = u_0$, $u'' = u' u_0' u$, $v_0'' = v_0$, and $v'' = v v_0' v'$.
		Observe that since $x \equiv_L y$ and thanks to the choice of $u, v, u_0, v_0 \in \widehat{\Sigma}^*$ given above, we have
		$u_0'' x^{\first} \in \widehat{\Sigma}^*_{\popeq}$,
		$u_0'' y^{\first} \in \widehat{\Sigma}^*_{\popeq}$,
		$x^{\last} v_0'' \in \widehat{\Sigma}^*_{\pusheq}$,
		$y^{\last} v_0'' \in \widehat{\Sigma}^*_{\pusheq}$,
		$\chain{u''}{u_0'' x v_0''}{v''}$, and
		$\chain{u''}{u_0'' y v_0''}{v''}$.
		Moreover, $u'' u_0'' x v_0'' v'' \in L $ iff $u'' u_0'' y v_0'' v'' \in L$, which is the same as $u' u_0' u u_0 x v_0 v v_0' v' \in L $ iff $u' u_0' u u_0 y v_0 v v_0' v' \in L$.
		Then, $u u_0 x v_0 v \sim u u_0 y v_0 v$, and thus $u u_0 x v_0 v \equiv_{L} u u_0 y v_0 v$.
		Therefore, ${\equiv_L}$ is chain-monotonic.
	\end{proof}

	\paragraph*{Lemma~\ref{lem:finite}}
	
	\proofsubparagraph*{Statement.}
	For every OPA $\A$ with $n$ states and $m$ input letters, the structural congruence $\equiv_{\A}$ has at most 
	$\mathcal{O}(m)^{\mathcal{O}(m \times n)^{\mathcal{O}(1)}}$ equivalence classes.
	\begin{proof}
		Suppose that $L$ is an OPL over $\widehat{\Sigma}$, and let $\A = (Q, \{q_I\}, F, \Delta)$ be a complete deterministic OPA with the unique initial state $q_I$ such that $L(\A) = L$.
		For every $w \in \widehat{\Sigma}^*$ the functions $f_w$ and $g_w$ have a finite input domain and a finite output range. 
		The functions $\varPhi_w$ however, have a finite input domain but an infinite output range. 
		Nevertheless, only the top of the output stack of $\varPhi_w$ is used in $\equiv_{\A}$ and, for all $w\in\widehat{\Sigma}^*$, the functions $(q, \gamma) \mapsto (\varPhi(q, \gamma))^{\top}$ do have a finite output range.
		Then, it is easy to see that $\equiv_{\A}$ has finitely many equivalence classes, thanks to Proposition~\ref{proposition:chain:finite}.
		In fact, it has at most:
		\[ \big(|\widehat{\Sigma}|^{2} \times 2^{2|\widehat{\Sigma}|}\big)
		\times \big(2^{|Q|}\big)^{|Q| \times (|\Gamma| + 1)}
		\times \big(2^{|Q|}\big)^{|Q|^2 \times (|\Gamma| + 1)}
		\times \big(2^{|\Gamma|+1}\big)^{|Q| \times (|\Gamma| + 1)} \]
		equivalence classes.
		We recall that $\Gamma = \widehat{\Sigma} \times Q$.
		Hence, in Landau's notation we obtain $|{\equiv_{\A}}| \leq \mathcal{O}(|\widehat{\Sigma}|)^{\mathcal{O}(|\Sigma|\times|Q|)^{\mathcal{O}(1)}}$.
	\end{proof}

	\paragraph*{Lemma~\ref{lem:coarse}}

\proofsubparagraph*{Statement.}
For every OPA $\A$, the congruence $\equiv_{L(\A)}$ is coarser than the congruence $\equiv_{\A}$.

\begin{proof}
	We claim that every class of $\equiv_{\A}$ is contained in a class of $\equiv_L$, thus establishing that $\equiv_L$ also has finitely many equivalence classes.
	Consider $x, y, u, v, u_0, v_0 \in \widehat{\Sigma}^*$ such that $x \equiv_{\A} y$.
	Assume that $u_0x^{\first}\in \widehat{\Sigma}^*_{\popeq}$, $x^{\last}v_0\in \widehat{\Sigma}^*_{\pusheq}$, and $\chain{u}{u_0xv_0}{v}$ hold.
	Then, since $x \approx y$, we have that $u_0y^{\first}\in \widehat{\Sigma}^*_{\popeq}$, $y^{\last}v_0\in \widehat{\Sigma}^*_{\pusheq}$ and $\chain{u}{u_0yv_0}{v}$ hold for all $u, v, u_0, v_0 \in \widehat{\Sigma}^*$ as well.
	Next, we prove that all configurations $(t, v, \theta)$ where $t\in Q$ and $\theta\in \varPhi_u(q_I, \bot)$ reachable from $(q_I, uu_0xv_0v, \bot)$ is also reachable from $(q_I, uu_0yv_0v, \bot)$.
	As the role of $x$ and $y$ is symmetrical, it implies that $uu_0xv_0v\in L \Leftrightarrow uu_0yv_0v \in L$.
	There are four cases, depending on the precedences between $u_0^{\last} \popeq x^{\first}$ and  $x^{\last} \pusheq v_0^{\first}$.

	We have $u^{\last} \push u_0^{\first}$ by definition, and we consider only the case where $u_0^{\last} \pop x^{\first}$ and $x^{\last} \shift v_0^{\first}$, as the other can be tackled similarly.
	From $(q_I, uu_0xv_0v, \bot)$ all configurations that $\A$ can reach after reading $u$ are of the from 
	$(q_u, u_0xv_0v, \theta_u \cdot \bot)$ where $q_u \in f_u(q_I, \bot)$ and $\theta_u \in \varPhi_u(q_I, \bot)$,
	because $\varepsilon \push u^{\first}$.
	After reading $u_0$, $\A$ can reach configurations of the form $(q_{u_0}, xv_0v, \theta_{u_0} \cdot \theta_u \cdot \bot)$ where $q_{u_0}\in f_{u_0}(q_u, \bot)$ and $\theta_{u_0}\in \varPhi_{u_0}(q_u, \bot)$,
	due to $u^{\last}u_0\in\widehat{\Sigma}^*_{\pusheq}$.
	Observe that we have been able to abstract the stack $\theta_u \cdot \bot$ with $\bot$ thanks to $u^{\last} \push u_0^{\first}$.
	Then $\A$ must performs pop-transitions since $u_0x^{\first}\in\widehat{\Sigma}^*_{\popeq}$.
	After popping $\theta_{u_0}$, the reachable configurations are of the form $(p_{u_0}, xv_0v, \theta_u \cdot \bot)$ where $p_{u_0} \in g_{u_0}(q_u, q_{u_0}, \bot)$.
	Observe that, the stack is clear from the computation of $u_0$ due to $u_0^{\last} \pop x^{\first}$, in fact $\chain{u}{u_0}{x}$.
	All configurations that $\A$ can reach after reading $x$ are of the form $(q_{x}, v_0v, \theta_{x} \cdot \theta_u \cdot \bot)$ where $q_{x} \in f_{x}(p_{u_0},\bot)$ and $\theta_{x}\in\varPhi_{x}(p_{u_0},\bot)$,
	because $u^{\last}x \in \widehat{\Sigma}^*_{\pusheq}$.
	Once again, we have been able to abstract the stack $\theta_u \cdot \bot$ with $\bot$, this time it is thanks to $u^{\last} \push x^{\first}$.
	Now, as we are dealing with $x^{\last} \shift v_0^{\first}$, $\A$ must clear the stack from the computation of $x$ after reading $v_0$, and so the function $g_{x}$ must be called after reading $g_{v_0}$.
	We emphasize that the stack $\theta_{x} \cdot \theta_u \cdot \bot$ cannot be abstracted by $\bot$.
	However, since $x^{\last}v_0 \in\widehat{\Sigma}^*_{\pusheq}$, it can be abstracted by its top symbol, denoted by $\theta_{x}^{\top}$.
	Observe that, in the case where $x^{\last}\shift v_0^{\first}$, we must have $\theta_x\neq \bot$, and thus the top of $\theta_{x}$ indeed correspond to the top of $\theta_{x} \cdot \theta_u \cdot \bot$.
	Let $\theta'_x$ be defined by $\theta_x = \theta_x^{\top}\theta'_x$.
	After reading $v_0$, $\A$ can reach configurations of the from $(q_{v_0}, v, \theta_{v_0}\cdot \theta'_x \cdot \theta_u \cdot \bot)$.
	where $q_{v_0} \in f_{v_0}(q_x, \theta_x^{\top})$ and $\theta_{v_0} \in \varPhi_{v_0}(q_x, \theta_x^{\top})$, 
	because $x^{\last}v_0\in \widehat{\Sigma}^*_{\pusheq}$.
	It is worth noticing that the top of the stack $\theta_{x}^{\top}$ may have been modified while reading $v_0$, due to $x^{\last} \shift v_0^{\first}$.
	For instance, it is the case when $v_0$ is a single letter.
	Then $\A$ must performs pop-transitions since $v_0v^{\first} \in\widehat{\Sigma}^*_{\popeq}$.
	After popping $\theta_{v_0}$, the reachable configurations are of the form $(p_{v_0}, v, \theta'_x \cdot \theta_u \cdot \bot)$ where $p_{v_0} \in g_{v_0}(q_x, q_{v_0}, \theta_x^{\top})$.
	Observe that, the stack is clear from the computation of $v_0$ due to $v_0^{\last} \pop v^{\first}$.
	Moreover, $g_{v_0}$ pops the modified top of $\theta_x$.
	More pop-transitions must be performed since $xv^{\first} \in\widehat{\Sigma}^*_{\popeq}$. 
	After popping $\theta'_x$, the reachable configurations are of the form  $(p_{x}, v, \theta_u \cdot \bot)$ where $p_{x}=g_{x}(p_{u_0}, p_{v_0}, \bot)$.
	This time, the stack is clear from the computation of $x$ due to $x^{\last} \pop v^{\first}$.
	We emphasize that $g_{x}$ is used as it have been defined for.
	Indeed, $g_{x}$ takes as parameters the current state $p_{v_0}$ and the parameters on which $f_x$ have been previously called, i.e., $p_{u_0}$ and $\bot$.
	By taking $t = p_{x}$ and $\theta=\theta_u\cdot\bot$, we established that $(q_I, uu_0xv_0v, \bot) \tikzpath (t, v, \theta)$.
	As a direct consequence of $x \equiv_{\A} y$, making the above reasoning with $y$ instead of $x$ results dealing with identical intermediate sets of states and sets of stacks. 
	Hence, we proved that all configurations of the from $(t, v, \theta)$ where $t\in Q$ and $\theta\in \varPhi_u(q_I, \bot)$ reachable from $(q_I, uu_0xv_0v, \bot)$ is also reachable from $(q_I, uu_0yv_0v, \bot)$.	
\end{proof}

	\paragraph*{Lemma~\ref{lem:recognizability}}
	\proofsubparagraph*{Statement.}
		For every $L \subseteq \widehat{\Sigma}^*$, if $\equiv_L$ has finite index then $L$ is a $\widehat{\Sigma}$-OPL.

	\begin{proof}
		Consider a language $L \subseteq \widehat{\Sigma}^*$ such that $\equiv_L$ has finitely many equivalence classes.
		We construct a deterministic OPA that recognizes $L$ and whose states are based on the equivalence classes of ${\equiv_L}$.
		Given $w\in\widehat{\Sigma}^*$, we denote $[w]$ its equivalence class with respect to ${\equiv_L}$.
		We construct $\A = (Q, \{q_0\}, F, \Delta)$ with the set of states $Q = \{([u], [v]) \st u, v\in \widehat{\Sigma}^*\}$, the initial state $q_0 = ([\varepsilon], [\varepsilon])$, the set of accepting states $F = \{([\varepsilon], [w]) \st w\in L \}$, and the $\widehat{\Sigma}$-driven transition function $\Delta \colon Q \times \Sigma \times (\Gamma^+ \cup \{\bot\}) \rightarrow Q \times (\Sigma \cup \{\varepsilon\}) \times (\Gamma^+ \cup \{\bot\})$, where $\Gamma=\Sigma \times Q$, is defined as follows:
		$\Delta$ maps $(([u], [v]), a, \stack{b,([u'],[v'])}\theta)$ to $(([a],[\varepsilon]), \varepsilon,\stack{a, ([u], [v])}\stack{b,([u'],[v'])}\theta)$ if $b \push a$,
		it returns $(([uva],[\varepsilon]), \varepsilon,\stack{a,([u'],[v'])}\theta)$ if $b\shift a$,
		and $(([u'],[v'uv]), a,\theta)$ if $b \pop a$.
		
		\subparagraph*{\textcolor{lipicsGray}{\normalfont Invariants.}}
		We show that the automaton $\A$ satisfies the following invariants after reaching the state $([u], [v])\in Q$ from the initial state $q_0$:
		\begin{enumerate}
			\item The top of the stack is $u^{\last}$ if $u\neq \varepsilon$, and $\bot$ otherwise.
			\item All outgoing on $a\in\Sigma$ satisfies $va\in\widehat{\Sigma}^*_{\popeq}$ and $v\neq\varepsilon \Rightarrow v^{\last}\pop a$.
			\item $u \in \widehat{\Sigma}^*_{\shift}$.
			\item $u^{\last}v\in\widehat{\Sigma}^*_{\pusheq}$ and $v \neq \varepsilon \Rightarrow u^{\last} \push v^{\first}$.
		\end{enumerate}
		We prove all invariants together by induction on the length of the run from $q_0$ that reaches the state $([u], [v])\in Q$.
		The run starts in $q_0= ([\varepsilon], [\varepsilon])$ which trivially satisfies all invariants.
		By induction we assume the invariants to hold for all run of length $n>0$.
		Let $([u], [v])$ be the state reached after $n$ transitions, $a\in\Sigma$ be the incoming letter and $(b, ([u'], [v']))$ be the top of the stack.

		If $b \push a$, the automaton $\A$ performs a push-transition to reach $([a], [\varepsilon])$, which satisfies (1) by definition of push-transition.
		The invariants (2, 3, 4) hold trivially.
	
		If $b\shift a$, the automaton $\A$ performs a shift-transition to reach $([uva], [\varepsilon])$, which satisfies (1) by definition of the shift-transition.
		Since a shift-transition is triggered, the stack is not $\bot$.
		By the induction hypothesis, (1) ensures that $u^{\last}=b$ and (2) ensures that $va \in \widehat{\Sigma}^*_{\popeq}$.
		Consequently and because $b\shift a$, we have that $uva \in \widehat{\Sigma}^*_{\shift}$, i.e., (3) is preserved.
		The invariants (2, 4) hold trivially.	
		
		If $b \pop a$, the automaton $\A$ performs a pop-transition to reach $([u'], [v'uv])$.
		Since a pop-transition is triggered, the stack is not $\bot$.
		By the induction hypothesis, (1) ensures that $u^{\last}=b$, thus $u^{\last} \pop a$.
		In particular $u\neq\varepsilon$, and thus $u^{\first} \in \Sigma$.
		Additionally, the induction hypothesis gives us that $va \in\widehat{\Sigma}^*_{\popeq}$ and $v\neq\varepsilon \Rightarrow v^{\last}\pop a$ by (2).
		We have that $uva \in\widehat{\Sigma}^*_{\popeq}$ because $u^{\last} \pop a$, $u\in\widehat{\Sigma}^*_{\shift}$ and $\chain{u}{v}{a}$ from (2, 3, 4).
		Finally, (4) ensures that $u^{\last}v\in\widehat{\Sigma}^*_{\pusheq}$.
		
		Since $([u'], [v'])$ is on the stack, there exists a strict prefix of the current run that ends in $([u'], [v'])$ and such that popping $(b, ([u'], [v']))$ recovers its stack.
		By the induction hypothesis on such smaller run, (1) ensures that if $u'= \varepsilon$ then the stack is $\bot$ otherwise the top is $u'^{\last}$, and (3) ensures that $u'\in\widehat{\Sigma}^*_{\shift}$.
		So, (1) and (3) are directly preserved.
		Continuing the induction hypothesis,
		(2) ensures that $v'u^{\first}\in\widehat{\Sigma}^*_{\popeq}$ and (4) ensures that $u'^{\last}v'\in\widehat{\Sigma}^*_{\pusheq}$ and  $v'\neq\varepsilon\Rightarrow u'^{\last} \push v'^{\first}$.
		In the case where $u'\neq\varepsilon$ and $v'= \varepsilon$, we get that $v'uv\neq\varepsilon\Rightarrow u'^{\last} \push (v'uv)^{\first}$ since $u'^{\last} \push u^{\first}$ as the automaton $\A$ pushed $([u'], [v'])$ on the stack while reading $u^{\first}$.
		The other cases are immediate.
		Moreover, $u'^{\last}v'uv\in\widehat{\Sigma}^*_{\pusheq}$ comes as we established that $u'^{\last}v'\in\widehat{\Sigma}^*_{\pusheq}$, $u\in\widehat{\Sigma}^*_{\shift}$ , $u^{\last}v\in\widehat{\Sigma}^*_{\pusheq}$, and $u'^{\last} \push u^{\first}$.
		Hence (4) is preserved.
		For the invariant (2), we established that $uva\in\widehat{\Sigma}^*_{\popeq}$ and $v'u^{\first}\in\widehat{\Sigma}^*_{\popeq}$, which implies that $v'uva\in\widehat{\Sigma}^*_{\popeq}$.
		We also established that $v\neq\varepsilon \Rightarrow v^{\last} \pop a$ and $u^{\last} \pop a$, which implies that $(v'uv)^{\last} \pop a$.
		In particular, the invariant (2) is preserved.
		
		\subparagraph*{\textcolor{lipicsGray}{\normalfont Determinism.}}
		For all states $([a], [\varepsilon]), ([b], [\varepsilon])\in Q$ reachable in $\A$ with a push-transition,
		if $a \neq b$ then $a \not\approx b$ which implies that $a \not\equiv_L b$.
		Reciprocally, if $a \equiv_L b$ then $a \approx b$, which implies that $a=b$.
		
		For all states $([u_1], [v_1]), ([u_2], [v_2])\in Q$ and all $a\in\Sigma$, let $([u_1v_1a], [\varepsilon]), ([u_2v_2a], [\varepsilon])$ be two states reachable in $\A$ with a shift-transition.
		We show that, if $u_1\equiv_L u_2$ and $v_1\equiv_L v_2$ then $u_1v_2a \equiv_L u_2v_2a$. 
		If $v_1=\varepsilon$, then $v_2 \equiv_L v_1$ implies $v_2=\varepsilon$.
		Also $u_1 \equiv_L u_2$ implies $u_1a \equiv_L u_2a$ by chain-monotonicity of ${\equiv_L}$.
		Otherwise, if $v_1\neq \varepsilon$, then $v_2 \equiv_L v_1$ implies $v_2\neq\varepsilon$.
		Furthermore, $\chain{u_1}{v_1}{a}$ and $\chain{u_1}{v_2}{a}$, since $u_1 \approx u_2$ and the invariants (2) and (4) hold.
		In particular, $u_1^{\last}v_2a \in \widehat{\Sigma}^*_{\shift}$.
		By chain-monotonicity of ${\equiv_L}$, we have $v_1 \equiv_L v_2$ implies $u_1v_1a \equiv_L u_1v_2a$, and $u_1 \equiv_L u_2$ implies $u_1v_2a \equiv_L u_2v_2a$
		Hence, $u_1v_1a \equiv_L u_2v_2a$, by transitivity of ${\equiv_L}$.
		
		For all states $([u_1], [v_1])$, $([u_2], [v_2])$, $([u'_1], [v'_1])$, $([u'_2], [v'_2])\in Q$, we let $([u'_1], [v'_1u_1v_1])$ and $([u'_2], [v'_2u_2v_2])$ be two states reachable in $\A$ with a pop-transition.
		We show that, if $u_1\equiv_L u_2$, $v_1\equiv_L v_2$, and $v'_1\equiv_L v'_2$, then $v'_1u_1v_1 \equiv_L v'_2u_2v_2$. 
		As a direct consequence of the invariants, we have that $\chain{v'_1u_1}{u_1}{\varepsilon}$, $\chain{v'_1u_1}{u_2}{\varepsilon}$, $\chain{\varepsilon}{v'_1u_1v_2}{\varepsilon}$, $\chain{\varepsilon}{v'_1u_2v_2}{\varepsilon}$, $\chain{\varepsilon}{v'_1}{u_2v_2}$, and $\chain{\varepsilon}{v'_2}{u_2v_2}$.
		Additionally, $v'_1u^{\first} , v'_1u_2^{\first}\in \widehat{\Sigma}^*_{\popeq}$ and $u_1^{\last}v_2, u_2^{\last}v_2 \in \widehat{\Sigma}^*_{\pusheq}$.
		By chain-monotonicity of ${\equiv_L}$, we have $v_1 \equiv_L v_2$ implies $v'_1u_1v_1 \equiv_L v'_1u_1v_2$, ${\equiv_L}$, $u_1 \equiv_L u_2$ implies $v'_1u_1v_2 \equiv_L v'_1u_2v_2$, and $v'_1 \equiv_L v'_2$ implies $v'_1u_2v_2 \equiv_L v'_2u_2v_2$.
		Hence, $v'_1u_1v_1 \equiv_L v'_2u_2v_2$, by transitivity of ${\equiv_L}$.
		
		\subparagraph*{\textcolor{lipicsGray}{\normalfont Correctness.}}
		We have $L(\A) = L$ since $[w] \cap L = \varnothing$ or $[w] \subseteq L$, for all $w\in\Sigma^*$.
		More precisely, every $w\in\widehat{\Sigma}^*$ admits a unique run $(([\varepsilon], [\varepsilon]), w, \bot)\tikzpath (([\varepsilon],[w']), \varepsilon, \bot)$ since $\A$ is deterministic.
		By induction we prove that, for all $x,y \in \Sigma^*$, all $\theta\in\Gamma^+\cup\{\bot\}$, if $(q_0, xy, \bot) \tikzpath (([u_0],[v_0], y, \theta)$ then there exist $n\in\N$, and $(a_i, ([u_i],[v_i]))_{i\in\{1\dots n\}}$ such that $\theta=\stack{a_1, ([u_1],[v_1])}...\stack{a_n,([u_n],[v_n])}\bot$ and $x \equiv_L u_nv_n\dots u_0v_0$.
		In particular, we get that $w \equiv_L w'$ implying that $w\in L(\A)$ iff $([\varepsilon], [w'])\in F$ iff $w'\in L$ iff $w \in L$.
		The base case, when the run has length $n=0$, is trivial since $x=\varepsilon$ and $\theta=\bot$.
		Suppose that the property holds for all runs of length $n$, we prove that it holds for runs of length $n+1$.
		Let $z=u_nv_n\dots u_0v_0$.
		In the case where the last transition is a push-transition that reads $a\in\widehat{\Sigma}$ from $([u_0], [v_0])$.
		If $z=\varepsilon$ then $x=\varepsilon$ since $x\equiv_L z$.
		Hence $xa \equiv_L za$ holds trivially.
		Otherwise $z\neq \varepsilon$.
		Since a push-transition is triggered and $u_0^{\last}v_0 \in \widehat{\Sigma}^*_{\pusheq}$ by invariant (2), we have that $\chain{\varepsilon}{z}{a}$.
		Since $x \equiv_{L} z$ and $\chain{\varepsilon}{x}{a}$ holds as well then we get $xa \equiv_L za$ by chain-monotonicity.
		In the case where the last transition is a pop-transition from $([u_0], [v_0])$.
		Then the reached state is $([u_1], [v_1u_0v_0])$ and the property is trivially preserved.
		In the case where the last transition is a shift-transition that reads $a\in\widehat{\Sigma}$ from $([u_0], [v_0])$.
		Since a shift-transition is triggered, we that the $u_0^{\last} \shift a$ by invariant (1).
		In particular $u_0\neq\varepsilon$.
		By invariant (2), if $v_0\neq\varepsilon$ then $v_0^{\last} \pop a$.
		Due to $x \equiv_{L} z$, we have that $x^{\last} = v_0^{\last}$.
		So, $xa \equiv_L za$ since $\chain{\varepsilon}{x}{a}$ and $\chain{\varepsilon}{z}{a}$.
		Otherwise if $v_0=\varepsilon$.
		Due to $x \equiv_{L} z$ and $u_0\neq\varepsilon$, we have that $x^{\last} = u_0^{\last}$.
		So, $xa \equiv_L za$ since $\chain{\varepsilon}{x}{a}$ and $\chain{\varepsilon}{z}{a}$.
	\end{proof}

	\paragraph*{Proposition~\ref{proposition:monotonicity}}
	\proofsubparagraph*{Statement.}
		For every OPA $\A$, $\leqslant_\A$ is a computable chain-monotonic well quasi-order.

	\begin{proof}
		Let $\widehat{\Sigma}$ be an operator precedence alphabet, and $\A = (Q, I, F, \Delta)$ be an OPA.
		We only show that the structural quasi-order $\leqslant_\A$ is chain-monotonic since the rest is argued before the statement in the main body of the paper.
		Let us define the following relation over $\widehat{\Sigma}^*$:
		\[x \ll y \iff
		\forall q \in Q, \forall \gamma \in \Gamma \cup \{\bot\}
		\bigwedge \begin{cases}
			f_{x}(q,\gamma) \subseteq f_{y}(q,\gamma) \\
			g_{x}(q,q',\gamma) \subseteq g_{y}(q,q',\gamma) \\ 
			(\varPhi_{x}(q,\gamma))^\top \subseteq (\varPhi_{y}(q,\gamma))^\top
		\end{cases}
		\]
		Recall that for every $x, y \in \widehat{\Sigma}^*$ we have $x \leqslant_\A y$ iff $x \approx y$ and $x \ll y$, where ${\approx}$ is the chain equivalence.
		In particular, we want to show that for every $x,y, u, v, u_0, v_0 \in \widehat{\Sigma}^*$ such that $u_0 x^{\first}, u_0 y^{\first} \in \widehat{\Sigma}^*_{\popeq}$, $x^{\last} v_0, y^{\last} v_0 \in \widehat{\Sigma}^*_{\pusheq}$, $\chain{u}{u_0 x v_0}{v}$, $\chain{u}{u_0 y v_0}{v}$, and $x \leqslant_{\A} y$, we have $u u_0 x v_0 v \leqslant_{\A} u u_0 y v_0 v$.
		Since the chain equivalence ${\approx}$ is chain-monotonic, we only need to show that ${\ll}$ is chain-monotonic, i.e., we have $u u_0 x v_0 v \ll u u_0 y v_0 v$ for every $x,y, u, v, u_0, v_0 \in \widehat{\Sigma}^*$ as above.
		
		Let $\star \notin \Sigma$ be a fresh letter for which we extend the precedence relation with $a \push \star$ for all $a \in \Sigma$.
		Let $w\in \widehat{\Sigma}^*$, $q, q' \in Q$, and $\gamma \in \Gamma \cup \{\bot\}$.
		Recall the following:
%		\begin{itemize}
%			\item $f_w(q, \gamma) = \{q_w \in Q \st \exists \gamma_w \in \Gamma^+\cup\{\bot\}, (q, w\star, \gamma) \tikzpath (q_w, \star,\gamma_w) \}$
%			\item $\varPhi_w(q, \gamma) = \{\gamma_w \in \Gamma^+\cup\{\bot\} \st \exists q_w \in Q, (q, w\star, \gamma) \tikzpath (q_w, \star,\gamma_w) \}$
%			\item $g_w(q, q', \gamma) = \{ p_w \in Q \st \exists \gamma_w \in \varPhi_w(q, \gamma), (q', \varepsilon, \gamma_w) \tikzpath (p_w, \varepsilon, \bot) \}$
%		\end{itemize}
		\[f_w(q, \gamma) = \{q_w \in Q \st \exists \gamma_w \in \Gamma^+\cup\{\bot\}, (q, w\star, \gamma) \tikzpath (q_w, \star,\gamma_w) \}\]
		\[\varPhi_w(q, \gamma) = \{\gamma_w \in \Gamma^+\cup\{\bot\} \st \exists q_w \in Q, (q, w\star, \gamma) \tikzpath (q_w, \star,\gamma_w) \}\]
		\[g_w(q, q', \gamma) = \{ p_w \in Q \st \exists \gamma_w \in \varPhi_w(q, \gamma), (q', \varepsilon, \gamma_w) \tikzpath (p_w, \varepsilon, \bot) \}\]
		Now, let $x,y, u, v, u_0, v_0 \in \widehat{\Sigma}^*$ such that $u_0 x^{\first}, u_0 y^{\first} \in \widehat{\Sigma}^*_{\popeq}$, $x^{\last} v_0, y^{\last} v_0 \in \widehat{\Sigma}^*_{\pusheq}$, $\chain{u}{u_0 x v_0}{v}$, $\chain{u}{u_0 y v_0}{v}$, and $x \leqslant_{\A} y$.
		Since $x \leqslant_{\A} y$ implies $x \approx y$, which implies $x^{\first} = y^{\first}$ and $x^{\last} = y^{\last}$, it is clear that from the definitions that $f_{u u_0 x v_0 v}(q, \gamma) \subseteq f_{u u_0 y v_0 v}(q, \gamma)$ and $\varPhi_{u u_0 x v_0 v}(q,\gamma) \subseteq \varPhi_{u u_0 y v_0 v}(q,\gamma)$
	%TODO: CHECK
%		$(\varPhi_{u u_0 x v_0 v}(q,\gamma))^\top \subseteq (\varPhi_{u u_0 y v_0 v}(q,\gamma))^\top$
%  OK: CONSEQUENCE OF THE INCLUSION OF STACKS
		for all $q \in Q$ and $\gamma \in \Gamma \cup \{\bot\}$.
		Intuitively, the reasoning for this is as follows:
		$\A$ reaches a set of configurations after processing $u u_0$.
		For every configuration in this set, consider the state and the top stack symbol given as inputs to $f_x$ and $f_y$, as well as $\varPhi_{x}$ and $\varPhi_{y}$, for whose outputs we know the inclusion relation above.
		It means that after reading $u u_0 x$ and $u u_0 y$ from any state and top stack symbol we have the same relations as well.
		Now, consider this time the states and top stack symbols of the configurations reached after $u u_0 x$ and $u u_0 y$.
		Proceeding the computation from these configurations with the suffix $v_0 v$ clearly preserves the inclusion relation for the sets of states reached.
		For the sets of stacks, note that $\chain{u}{u_0 x v_0}{v}$ and $\chain{u}{u_0 y v_0}{v}$, therefore the suffix $v_0 v$ pops the stack beyond what has been pushed while reading $x$ and $y$, which is the same for both words.
		%Thanks to this stronger relation between the sets of stacks reached after reading $u u_0 x v_0 v$ and $ u u_0 y v_0 v$, it is also easy to see that $g_{u u_0 x v_0 v}(q, q', \gamma) \subseteq g_{u u_0 y v_0 v}(q, q', \gamma)$ for every $q,q' \in Q$ and $\gamma \in \Gamma \cup \{\bot\}$.
		Finally,  for all  $q,q' \in Q$ and all $\gamma \in \Gamma \cup \{\bot\}$, we have that $g_{u u_0 x v_0 v}(q, q', \gamma) \subseteq g_{u u_0 y v_0 v}(q, q', \gamma)$ from $\varPhi_{u u_0 x v_0 v}(q,\gamma) \subseteq \varPhi_{u u_0 y v_0 v}(q,\gamma)$ which implies that $\{(q', \varepsilon, \gamma_w) \st \gamma_w\in \varPhi_{u u_0 x v_0 v}(q,\gamma)\} \subseteq \{(q', \varepsilon, \gamma_w) \st \gamma_w\in  \varPhi_{u u_0 y v_0 v}(q,\gamma)\}$.
		%the starting configurations for defining the set $g_{u u_0 x v_0 v}$ is a subset of those of $g_{u u_0 y v_0 v}$.
		Therefore, $u u_0 x v_0 v \ll u u_0 y v_0 v$, and thus $u u_0 x v_0 v \leqslant_\A u u_0 y v_0 v$, implying that $\leqslant_\A$ is chain-monotonic.	
	\end{proof}

	\paragraph*{Lemma~\ref{lemma:preserve}}
	\proofsubparagraph*{Statement.}
		For every OPA $\A$ and $w_1, w_2 \in \widehat{\Sigma}^*$, if $w_1 \leqslant_\A  w_2$ and $w_1 \in L(\A)$ then $w_2 \in L(\A)$.

	\begin{proof}
		Let $\A = (Q, I, F, \Delta)$.
		If $w_1 \in L(\A)$ then $(q_I, w_1, \bot) \tikzpath (q_F, \varepsilon, \bot)$ for some $q_I \in I$ and $q_F \in F$.
		Since $w_1 \leqslant_\A  w_2$, we also have that $(q_I, w_1, \bot) \tikzpath (q_F, \varepsilon, \bot)$ implying that $w_2\in L(\A)$.
	\end{proof}

	\paragraph*{Lemma~\ref{lemma:cat}}
	\proofsubparagraph*{Statement.}
	Let $\A=(Q, I, F, \Delta)$ be an OPA, and let $\Gamma=\Sigma \times Q$.
	Considering $\vec{U}^{a, b,c}_{s, t} = \varepsilon$ when $a=b$ and $s=t$, otherwise $\vec{U}^{a, b,c}_{s, t} = \varnothing$.
	The following holds for all $n>0$:
	\[\texttt{Cat}^n(\vec{U}^{a, b, c}_{s, t}) {=} \big\{ u \mid
	(s, uc, \alpha){\tikzpath}(t, c, \beta),
	|u|=n, 
	\alpha\in\varTheta_a, \beta\in\varTheta_b,
	au\in\widehat{\Sigma}^*_{\pusheq},
	uc\in\widehat{\Sigma}^*_{\popeq}, u^{\last}=b
	\big\}\]
	where, for all $a\in\widehat{\Sigma}$, the set of stack symbols $\varTheta_a \subseteq \Gamma \cup \{\bot\}$ is defined by $\varTheta_a=\{\bot\}$ if $a=\varepsilon$, and $\varTheta_a=\{\stack{a, q} \st q\in Q\}$ otherwise.
	\begin{proof}
		For readability, we define the set of runs $\varOmega^{a, b, c}_{s, t, n}(w)$, for all $a, b, c\in\widehat{\Sigma}$, $s, t\in Q$, $n\in\N$ and $w\in\widehat{\Sigma}^*$ as follows.
		\[\varOmega^{a, b, c}_{s, t, m}(w) = \left\{ 
			\begin{array}{c|c}
				\rho & \bigwedge
		\begin{array}{l}
			\rho=(s, wc, \alpha)\tikzpath[m] (t, c, \beta)\land m \leq 2|u|\\
			\alpha\in\varTheta_a \land \beta\in\varTheta_b\\
			aw\in\widehat{\Sigma}^*_{\pusheq} \land wc\in\widehat{\Sigma}^*_{\popeq} \land (aw)^{\last}=b
		\end{array}\end{array}
		\right\}\]
		We reformulate the statement as $u \in \texttt{Cat}^n(\vec{U}^{a, b, c}_{s, t})$ if and only if $\varOmega^{a, b, c}_{s, t, 2n}(u) \neq \varnothing$.
		For all $u \in \texttt{Cat}^n(\vec{U}^{a, b, c}_{s, t})$, it takes a simple induction on $|u|$ to prove that $\varOmega^{a,b,c}_{s,t,2n}(u) \neq \varnothing$, because $\texttt{Cat}$ follows runs of $\A$ and preserves the invariants
		$m\leq 2|u|$,
		$au\in\widehat{\Sigma}^*_{\pusheq}$,
		$uc\in\widehat{\Sigma}^*_{\popeq}$,
		$(au)^{\last}=b$ by definition.
		Next, we prove that $\varOmega^{a,b,c}_{s,t,2n}(u) \neq \varnothing$ implies $u \in \texttt{Cat}^{n}(\vec{U}^{a, b, c}_{s, t})$, where $n=|u|$.

		The proof goes by induction on the structure of $w = a u c$.
		In the base case, $w$ does not admit any subchains, i.e., $\lambda(w)=w$.
		If $\varOmega^{a, b, c}_{s, t, 2n}(u) \neq \varnothing$ holds, then, $au\in\widehat{\Sigma}^*_{\pusheq}$ and $uc\in \widehat{\Sigma}^*_{\popeq}$.
		Together with $\lambda(w)=w$, it implies that $u\in\widehat{\Sigma}^*_{\shift}$.
		Hence, any run witnessing $\varOmega^{a, b, c}_{s, t, 2n}(u) \neq \varnothing$ performs exclusively shift-transitions.
		Actually, since that run performs exclusively shift-transitions, it length must be $n=|u|$.
		The base case goes by induction on such a witnessing run of the length $n$.
		Having $n=0$ implies $u=\varepsilon$ and $(s, ua, \alpha) = (t, a, \beta)$.
		In fact $\vec{U}$ is defined such that $\varepsilon\in\vec{U}^{a, b, c}_{s, t, a}$ exactly when $a=b$ and $s=t$.
		By induction, we assume that there exists a run of $\varOmega^{a, b, c}_{s, t, 2n}(u)$ of length $n=|u|$ of the form $(s, uc, \alpha) \tikzshift{a'} (q, vc, \theta) \tikzpath[n-1] (t, c, \beta)$ where $u=a'v$ and $a'=\theta^{\top}$.
		In particular $(q, vc, \theta) \tikzpath[n-1] (t, c, \beta)$ witnesses $\varOmega^{a', b, c}_{q, t, 2|v|}(v) \neq \varnothing$.
		So, $v \in \texttt{Cat}^{|v|}(\vec{U}^{a', b, c}_{q, t})$ by induction hypothesis.
		Finally, $u\in\texttt{Cat}^n(\vec{U}^{a, b, c}_{s, t})$ by definition of $\texttt{CatShift}$.
		
		In the inductive step, we assume that $w$ is of the form $a_0 u_0 a_1 u_1 \ldots a_k u_k a_{k+1}$ such that for all $0 \leq i \leq k$, either $\chain{a_i}{u_i}{a_{i+1}}$ or $u_i = \varepsilon$, and $\lambda(a_0a_1 \dots a_{k+1})=a_0a_1 \dots a_{k+1}$.
		It is worth emphasizing that $a_0=a$, and $b=(au)^{\last}$ and $a_{k+1}=c$ by definition of $w$.
		Also $k>0$, since otherwise $\lambda(w)=w$, which is the base case of this induction. 
		As in the base case, if $\varOmega^{a, b, c}_{s, t, 2n}(u) \neq \varnothing$ then $au\in\widehat{\Sigma}^*_{\pusheq}$ and $uc\in \widehat{\Sigma}^*_{\popeq}$, which implies $a_0a_1\dots a_{k+1}\in\widehat{\Sigma}^*_{\shift}$ since $\lambda(a_0a_1 \dots a_{k+1})=a_0a_1 \dots a_{k+1}$.
		If there exists a run $\rho$ witnessing $\varOmega^{a_0,(au)^{\last},a_{k+1}}_{s,t,2n}(u) \neq \varnothing$ with $n=|u|$ then, due to 
		$a_0a_1\dots a_{k+1}\in\widehat{\Sigma}^*_{\shift}$, for all $1 \leq i \leq k$, there exists a run $\rho_i$ over $u_i$ such that $\rho = \rho_0 \tikzshift{a_1} \rho_1 \tikzshift{a_2} \dots  \tikzshift{a_k} \rho_k$.
		Since $\chain{a_i}{u_i}{a_{i+1}}$ for all $0 \leq i \leq k$, each $\rho_i$ witnesses $\varOmega^{a_i, (a_iu_i)^{\last}, a_{i+1}}_{s_i, t_i, 2n_i}(u_i)\neq\varnothing$ with $n_i=|u_i|$ and $s_i, t_i\in Q$.
		However, the induction hypothesis cannot be apply on $\varOmega^{a_i, (a_iu_i)^{\last}, a_{i+1}}_{s_i, t_i, 2n_i}(u_i)\neq\varnothing$, as it may have the same chain structure as $w$.
		Hence, we proceed by cases to prove that $u_i \in \texttt{Cat}^{n_i}(\vec{U}^{a_i, (a_iu_i)^{\last}, a_{i+1}}_{s_i, t_i})$.
		In the case when $u_i=\varepsilon$ the run  $\rho_i$ witnessing $\varOmega^{a_i, (a_iu_i)^{\last}, a_{i+1}}_{s_i, t_i, 2n_i}(u_i)\neq\varnothing$, or equivalently $\varOmega^{a_i, a_i, a_{i+1}}_{s_i, t_i, 2n_i}(u_i)\neq\varnothing$ must be empty since $a_i \shift a_{i+1}$.
		This implies that $s_i=t_i$.
		In fact, we trivially have that $\varepsilon \in \vec{U}^{a_i, a_i, a_{i+1}}_{s_i, s_i, 2n_i}(u_i)$ by definition of $\vec{U}$.
		Equivalently, $u_i \in \texttt{Cat}^{n_i}\vec{U}^{a_i, (a_iu_i)^{\last}, a_{i+1}}_{s_i, t_i, 2n_i}(u_i)\neq\varnothing$.
		Otherwise $u_i\neq\varepsilon$.
		Let $b'_i=u_i^{\first}$, $b_i=u_i^{\last}$, and $u'_i$ be such that $u_i = b'_iu'_i$.
		In this cases, the run $\rho_i$ witnessing $\varOmega^{a_i, (a_iu_i)^{\last}, a_{i+1}}_{s_i, t_i, 2n_i}(u_i)\neq\varnothing$ must starts with a push-transition on $b'_i$ since $a_i \push b'_i$, and must ends with a pop-transition on $b_i$ since $b_i \pop a_{i+1}$.
		In other words, there exists some run $\rho'_i$ witnessing $\varOmega^{b'_i, b_i, a_{i+1}}_{s'_i, t'_i, 2n'_i}(u'_i)\neq\varnothing$ where $|u'_i|=n'_i$ and $s'_i, t'_i\in Q$.
		Now, we can apply the induction hypothesis on $w'=b'_iu'_ia_{i+1}$ which is a strict subchain of $w$.
		Hence, $u'_i \in \texttt{Cat}^{n'_i}(\vec{U}^{b'_i, b_i, a_{i+1}}_{s'_i, t'_i})$.
		The rest of the proof is straightforward.
		For all $1 \leq i \leq k$, we get $u_i \in \texttt{Cat}^{n_i}\vec{U}^{a_i, (a_iu_i)^{\last}, a_{i+1}}_{s_i, t_i, 2n_i}(u_i)\neq\varnothing$ by definition of $\texttt{CatChain}$ and since $\chain{a_i}{u_i}{a_{i+1}}$.
		Finally, we can prove that $u\in\texttt{Cat}^n(\vec{U}^{a, b, c}_{s, t})$ by induction on $k$, by definition of $\texttt{CatShift}$ and since $a_0a_1\dots a_{k+1}\in\widehat{\Sigma}^*_{\shift}$.
	\end{proof}
	
	\paragraph*{Lemma~\ref{lemma:base}}
	\proofsubparagraph*{Statement.}
		Let $\preccurlyeq$ be a chain-monotonic quasi-order over $\widehat{\Sigma}^*$.
		For every $A = (Q, I, F, \Delta)$ and $\vec{X}, \vec{Y}$ such that $\mathfrak{B}(\vec{X}^{a, b, c}_{s, t} \preccurlyeq \vec{Y}^{a, b, c}_{s, t})$ holds for all $s, t \in Q$ and all $a, b, c\in\Sigma \cup \{\varepsilon\}$, we have $\mathfrak{B}(\texttt{Cat}(\vec{X}^{a, b, c}_{s, t}) \preccurlyeq \texttt{Cat}(\vec{Y}^{a, b, c}_{s, t}))$ holds also for all $s, t \in Q$ and all $a, b, c\in\Sigma \cup \{\varepsilon\}$.

	\begin{proof}
		Assume that $\mathfrak{B}(\vec{X}^{a, b, c}_{s, t} \preccurlyeq \vec{Y}^{a, b, c}_{s, t})$ holds for all $s, t \in Q$ and all $a, b, c\in\Sigma\cup\{\varepsilon\}$.
		In particular, for all $y_0 \in \vec{Y}^{a, b, c}_{s, t}$, there exists $x_0 \in \vec{X}^{a, b, c}_{s, t}$ such that $x_0 \preccurlyeq y_0$.
		Consider $y \in \texttt{Cat}(\vec{Y}^{a, b, c}_{s, t})$, we show that there exists $x\in \texttt{Cat}(\vec{X}^{a, b, c}_{s, t})$ such that $x \preccurlyeq y$.
		By definition of $\texttt{Cat}$, there are three cases: Either (1) $y\in\vec{Y}^{a, b, c}_{s, t}$ or, (2) $y \in \texttt{CatShift}(\vec{Y}^{a, b, c}_{s, t})$, or (3) $y \in \texttt{CatChain}(\vec{Y}^{a, b, c}_{s, t})$.
		We show (2) since (3) can be prove similarly and (1) is trivial from $\mathfrak{B}(\vec{X}^{a, b, c}_{s, t} \preccurlyeq \vec{Y}^{a, b, c}_{s, t})$.
		Suppose that $y$ is of the form $y_1b'y_2$ for some $y_1 \in \vec{Y}^{a, a', b'}_{s, s'}$, $b'\in \Sigma$, and $y_2\in \vec{Y}^{b', b, c}_{t', t}$.
		By hypothesis, there exist $x_1\in \vec{X}^{a, a', b'}_{s, s'}$ and $x_2\in \vec{X}^{b', b, c}_{t', t}$ such that $x_1 \preccurlyeq y_1$ and $x_2\preccurlyeq y_2$.
		If $y_1=\varepsilon$ then $x_1=\varepsilon$ and thus $x_1b' \preccurlyeq y_1b'$.
		If $y_1\neq\varepsilon$ then $y_1^{\last} = a' \pusheq b'$ and $x_1^{\last} \pusheq b'$ since $y_1 \approx x_1$.
		We have that $\chain{\varepsilon}{x_1b'}{\varepsilon}$ and $\chain{\varepsilon}{y_1b'}{\varepsilon}$.
		So, $x_1b' \preccurlyeq y_1b'$.
		We have that $\chain{\varepsilon}{x_1b'y_2}{\varepsilon}$ and $\chain{\varepsilon}{y_1b'y_2}{\varepsilon}$.
		So, $x_1b'y_2 \preccurlyeq y_1b'y_2$.
		If $b' \push y_2$ then $b' \push x_2$.
		We have that $\chain{x_1b'}{x_2}{\varepsilon}$ and $\chain{x_1b'}{y_2}{\varepsilon}$.
		So, $x_1b'x_2 \preccurlyeq x_1b'y_2$.
		By transitivity, $x_1b'x_1 \preccurlyeq x_1b'y_2$.
		If $b' \shift y_2$ then $b' \shift x_2$.
		We have that $\chain{\varepsilon}{x_1b'x_2}{\varepsilon}$ and $\chain{\varepsilon}{x_1b'y_2}{\varepsilon}$.
		So, $x_1b'x_2 \preccurlyeq x_1b'y_2$.
		By transitivity, $x_1b'x_1 \preccurlyeq y_1b'y_2$.
	\end{proof}

	\paragraph*{Theorem~\ref{theorem:correct}}
	\proofsubparagraph*{Statement.}
		The algorithm from \figurename~\ref{algo} terminates and decides language inclusion.

	\begin{proof}
		First, we show that the inclusion algorithm from \figurename~\ref{algo} always terminates.
		From the definition of $\texttt{Cat}$ and the constant $\vec{U}$, we have that each component of $\vec{X}$ holds a finite set of words after executing finitely many instructions.
		The halting conditions of the repeat/until loop is effectively computable.
		Indeed, deciding $\mathfrak{B}(X \preccurlyeq Y)$ where $X$, $Y$ are finite sets and $\preccurlyeq$ a decidable quasi-order, can be done by checking whether $X \subseteq Y$ and that for every $y \in Y$ there exists $x \in X$ such that $x \preccurlyeq y$.
		Additionally, the quasi-order $\preccurlyeq$ is a well-quasiorders and thus, there is no infinite sequence $\{X_i\}_{i\in\N}$ such that ${_\preccurlyeq}{\upharpoonleft} X_1 \subsetneq {_\preccurlyeq}{\upharpoonleft} X_2 \subsetneq \dots$. 
		Since $\mathfrak{B}(X \preccurlyeq Y)$ is defined by $X \subseteq Y \land {_\preccurlyeq}{\upharpoonleft} X = {_\preccurlyeq}{\upharpoonleft} Y$ and since  $\texttt{Cat}$ only extends the upward closures of the components of $\vec{X}$, we find that the repeat/until loop must terminate after finitely many iterations.
		
		Now, we show that the inclusion algorithm from \figurename~\ref{algo} is correct.
		Supposing that the algorithm returns $\texttt{ko}$.
		Once $\vec{X}$ reached the fixpoint computed by the repeat/until loop, all $w \in \vec{X}^{\varepsilon, \varepsilon, \varepsilon}_{q_I, q_F}$ where $q_I \in I$ and $q_F \in F$ belong to $L_1$ by Lemma~\ref{lemma:cat}.
		Hence, when the algorithm returns $\texttt{ko}$, then $L_1 \not\subseteq L_2$.
		Conversely, supposing that $L_1 \not\subseteq L_2$, in particular let $w \in L_2 \setminus L_1$.
		In fact, $w$ belongs to $\texttt{Cat}^n(\vec{U}_{q_I, q_F}^{\varepsilon, \varepsilon, \varepsilon})$ for some $q_I \in I$, $q_F\in F$ and $n \in \N$, by Lemma~\ref{lemma:cat}.
		Additionally, observe that once $\vec{X}$ reached the fixpoint computed by the repeat/until loop, $\vec{X}_{q_I, q_F}^{\varepsilon, \varepsilon, \varepsilon}$ is a base of  $\texttt{Cat}^n(\vec{U}_{q_I, q_F}^{\varepsilon, \varepsilon, \varepsilon})$, i.e.\ $\mathfrak{B}(\vec{X}_{q_I, q_F}^{\varepsilon, \varepsilon, \varepsilon} \preccurlyeq \texttt{Cat}^n(\vec{U}_{q_I, q_F}^{\varepsilon, \varepsilon, \varepsilon}))$.
		This can be proved by induction thanks to Lemma~\ref{lemma:base}.
		Hence, there exists $w_0\in \vec{X}_{q_I, q_F}^{\varepsilon, \varepsilon, \varepsilon}$ such that $w_0 \preccurlyeq w$.
		Since $\preccurlyeq$ satisfies that $(w_0\preccurlyeq w \land w_0\in L_2) \implies w\in L_2$, we get $w_0 \notin L_2$ and thus the algorithm returns $\texttt{ko}$.
	\end{proof}

\end{document}